# Development of a $(4 - \epsilon)$-dimensional theory for the density of states of a disordered system near the Anderson transition

I M Suslov

## Contents

**Abstract.** The calculation of the density of states for the Schrödinger equation with a Gaussian random potential is equivalent to the problem of a second-order transition with a 'wrong' sign for the coefficient of the quartic term in the Ginzburg–Landau Hamiltonian. The special role of the dimension $d = 4$ for such a Hamiltonian can be seen from different viewpoints but is fundamentally determined by the renormalizability of the theory. The construction of an $\epsilon$ expansion in direct analogy with the phase-transition theory gives rise to the problem of a 'spurious' pole. To solve this problem, a proper treatment of the factorial divergency of the perturbation series is necessary. Simplifications arising in high dimensions can be used for the development of a $(4 - \epsilon)$-dimensional theory, but this requires successive consideration of four types of theories: a nonrenormalizable theories for $d > 4$, nonrenormalizable and renormalizable theories in the logarithmic situation ($d = 4$), and a super-renormalizable theories for $d < 4$. An approximation is found for each type of theory giving asymptotically exact results. In the $(4 - \epsilon)$-dimensional theory, the terms of leading order in $1/\epsilon$ are only retained for $N \sim 1$ ($N$ is the order of the perturbation theory) while all degrees of $1/\epsilon$ are essential for large $N$ in view of the fast growth of their coefficients. The latter are calculated in the leading order in $N$ from the Callan– Symanzik equation with the results of Lipatov method used as boundary conditions. The qualitative effect is the same in all four cases and consists in a shifting of the phase transition point in the complex plane. This results in the elimination of the 'spurious' pole and in regularity of the density of states for all energies. A discussion is given of the calculation of high orders of perturbation theory and a perspective of the $\epsilon$ expansion for the problem of conductivity near the Anderson transition.

## 1. Introduction

The contemporary theory of disordered systems [1–7] originated in the pioneering work of Anderson [1], who stated the possibility of quantum diffusion being destroyed by disorder. The model used by Anderson (and named after him) is the Schrödinger equation in the tight-binding approximation

$$\sum_{x'} J_{x-x'} \Psi_{x'} + V_x \Psi_x = E \Psi_x \qquad (1.1)$$

on a $d$-dimensional cubic lattice (whose points are numbered by the vector subscripts $x$ and $x'$); the overlap integrals $J_{x-x'}$ fall off rapidly as $|x - x'|$ increases, and the values of the potential $V_x$ at lattice points are independent random variables with a distribution (of width $W$), which is usually assumed to be rectangular [1] or Gaussian:

$$P\{V_x\} \sim \exp\left(-\sum_x \frac{V_x^2}{2W^2}\right). \qquad (1.2)$$

The spectrum of an ideal lattice

$$\epsilon(p) = \sum_x J_x \exp(-\mathrm{i} p x) \qquad (1.3)$$

**I M Suslov** P L Kapitza Institute for Physical Problems, Russian Academy of Sciences, ul. Kosygina 2, 117334 Moscow, Russia
Tel. (7-095) 137 79 85
Fax (7-095) 938-20 30
E-mail: suslov@kapitza.ras.ru

corresponds to the existence of a band of finite width $J$. According to Anderson, there is no diffusion in the band when $W \gg J$ and, as $W$ decreases, diffusion starts at a certain critical value of the ratio $W/J$. Anderson's method consisted in constructing a perturbation-theory expansion in powers of $J_{x-x'}$ with a subsequent rough evaluation and a rather subjective selection of diagrams.

The Anderson model is a discrete version of the standard Schrödinger equation with a random potential $V(x)$. This becomes obvious if we rewrite Eqn (1.1) in the form

$$[\epsilon(\hat{p}) + V(x)]\Psi(x) = E\Psi(x), \quad (1.4)$$

observing that $\exp(i\hat{p}x)$ is the operator of displacement by vector $x$ (the origin of energy is selected in such a way that $\epsilon(p) = p^2/2m$ at $p \ll a_0^{-1}$, where $a_0$ is the lattice constant). This equation was studied in the early 1960s by I M Lifshitz [8], who demonstrated the existence of a macroscopically large number of states outside of the spectrum of the original band, localized on the fluctuations of a random potential. According to Mott [2], the localized states are separated from the extended states (i.e., those which spread throughout the system) by the critical energy $E_c$, known as the mobility edge, since there is no static conductivity over the localized states at $T = 0$. In the finite-width band there are two mobility edges corresponding to the upper and lower edges of the band. As the amplitude of disorder $W$ increases, these two values converge and meet at a certain critical value of $W_c$; in this way, the absence of diffusion discovered by Anderson is associated with the localization of all states in the band. The metal–insulator transition, which occurs when the Fermi level crosses $E_c$ as a result of a change in the concentration of carriers or the degree of disorder, is known as the Anderson transition. This transition is characterized by the singular behavior of the conductivity $\sigma$ and the localization radius of wave functions $\xi$, which are usually described by power laws

$$\sigma \sim |E_F - E_c|^s, \quad \xi \sim |E_c - E_F|^{-\nu}, \quad (1.5)$$

where the critical exponents $s$ and $\nu$ are introduced by analogy with the theory of phase transitions [9–11].

The qualitative theory formulated by Mott was published in the late 1960s, and immediately found great resonance. Anderson's paper [1] was thoroughly analyzed, not without misconceptions, of which one is directly related to the topic of our review. In everyone's opinion, Anderson's paper was 'poorly written and incomprehensible,' which stimulated J Ziman, who is known to be a gifted popularizer, to 'explain everything better' [12]. He retained the essentials of Anderson's diagram analysis, but made one 'minor' modification. To wit, Anderson had used the exact Green's functions of Eqn (1.1), which are expressed in terms of its eigenfunctions $\psi_s(\mathbf{r})$ and eigenvalues $\epsilon_s$ ($s = 1, 2, \ldots, N$),

$$G_E^{R,A}(\mathbf{r}, \mathbf{r}') = \sum_s \frac{\psi_s(\mathbf{r})\psi_s^*(\mathbf{r}')}{E - \epsilon_s \pm i\delta}, \quad (1.6)$$

and had based his estimates on the typical (most probable) values. By contrast, Ziman used the averaged Green's functions, for which the diagram technique had been developed by Edwards [13], Abrikosov and Gor'kov [14]. Such an approach was more customary for most theoreticians, and Ziman's interpretation was accepted. Shortly afterwards, however, was Lloyd's paper published [15], which gave the exact averaged Green's function for Eqn (1.1)

with an energy distribution at lattice points of the form

$$P(V_x) = \frac{1}{\pi} \frac{W}{V_x^2 + W^2}. \quad (1.7)$$

In the momentum representation, the retarded Green's function had a simple form

$$\langle G(p, E) \rangle = \frac{1}{E - \epsilon(p) + iW} \quad (1.8)$$

and did not exhibit any singularities with respect to the energy $E$ or the amplitude of the random potential $W$, thus casting doubts on Anderson's results. Anderson argued that his findings applied specifically to the typical (not the averaged) Green's function, and that in his paper of 1958 he had specially emphasized the necessity of distinguishing between typical and average values. The fact is that if a distribution $P(X)$ of a random quantity $X$ has a slow power-law tail at large $X$, then its average, as defined by the integral $\int_0^\infty XP(X)\,dX$, possibly has nothing in common with the typically observed values of $X$.

Economou and Cohen [16] later demonstrated that a treatment based on averaged values is possible but requires some caution. Indeed, the Green's function $G(x, x')$ defines the *amplitude* of the transition from point $x$ to point $x'$, whereas it is the *probability* of this transition that is relevant to the kinetics. The use of the mean $\langle |G(x,x')|^2 \rangle$ allows one to introduce a criterion of localization actually equivalent to the Anderson criterion, whereas the mean $\langle G(x,x') \rangle$ is not relevant to the kinetics.

The mean Green's function, however, defines the directly observable density of states $\nu(E)$, and the regularity of the latter at a point that is singular for conductivity is marvelous indeed. Today this phenomenon is commonly recognized, although remains unproved for the general case. Physically, this is explained [17] by the fact that the averaged Green's function $\langle G(x,x') \rangle$, which is a function of the difference in arguments $x - x'$, is proportional to $\exp(-|x-x'|/l)$ in the domain of low disorder and is a short-range function [14]. According to Mott [2], the free path $l$ also preserves its meaning in the domain of high disorder (although it cannot be calculated from the kinetic equation), and can only decrease as the disorder increases. Accordingly, $\langle G(x,x') \rangle$ is always a short-range function and does not 'feel' the transition to the thermodynamic limit, which alone gives rise to singularities of physical quantities† [9–11]. By contrast, the quantity

$$\phi(\mathbf{r}_1\mathbf{r}_2, \mathbf{r}_3\mathbf{r}_4) = \langle G_{E+\omega}^R(\mathbf{r}_1\mathbf{r}_2) G_E^A(\mathbf{r}_3\mathbf{r}_4) \rangle, \quad (1.9)$$

which contains all the information about the kinetic properties, is of a long-range nature, which is associated with the presence of diffusion poles in its Fourier transform [19].

We see that the calculation of the density of states and the calculation of the conductivity of a disordered system are two essentially different problems: the former requires calculating the mean Green's function $\langle G(x, x')\rangle$ defined by the diagram series (Fig. 1a), which is routinely reduced to the calculation of the self-energy $\Sigma$,

$$\langle G_E^{R,A}(\mathbf{k})\rangle \equiv G_\mathbf{k}^{R,A} = \frac{1}{E - \epsilon_\mathbf{k} - \Sigma_\mathbf{k}^{R,A}}, \quad (1.10)$$

for which the series only contains irreducible diagrams (Fig. 1b). At the same time, the kinetic properties of a disordered system are associated with the quantity $\phi$, which is determined by a set of four-end diagrams built on $G^R$ and $G^A$ lines (Fig. 2a), and whose properties are similar to those of the two-particle Green's function in the theory of interacting particles. It satisfies the Bethe–Salpeter equation (Fig. 2b), which contains an irreducible vertex $U$ (Fig. 2c) and in the momentum representation has the form

$$\phi_{\mathbf{k}\mathbf{k}'}(\mathbf{q}) = G_{\mathbf{k}+\mathbf{q}/2}^R G_{\mathbf{k}-\mathbf{q}/2}^A \left[ N\delta_{\mathbf{k}-\mathbf{k}'} + \frac{1}{N}\sum_{\mathbf{k}_1} U_{\mathbf{k}\mathbf{k}_1}(\mathbf{q})\phi_{\mathbf{k}_1\mathbf{k}'}(\mathbf{q}) \right]. \quad (1.11)$$

---

† On closer consideration, the argument with Lloyd's model does not seem convincing. Mathematically, Lloyd's model is highly degenerate: it is based on an esoteric diagram technique (in which all diagrams with intersections of impurity lines vanish), and application of the replica method reduces the model to a Gaussian functional integral [cf. Eqn (1.14) below]. There is no evidence that the specific features of Lloyd's model are related to the physical content of the problem — on the contrary, for hierarchical models [18] it is possible to demonstrate explicitly that the potentials with infinite dispersion belong to a different universality class.

Here we assume that the energy variable is $E + \omega$ for $G^R$ and $E$ for $G^A$, and use the three-momentum notation (Fig. 2d).

Actually, these two problems are not quite independent. The analysis in the parquet approximation reveals [5, 20] that the mathematical difficulties in both cases are of the same nature, and are associated with the problem of the spurious pole (Section 3). On the other hand, there is Ward's identity [21]

$$\Delta\Sigma_\mathbf{k}(\mathbf{q}) = \frac{1}{N}\sum_{\mathbf{k}'} U_{\mathbf{k}\mathbf{k}'}(\mathbf{q})\Delta G_{\mathbf{k}'}(\mathbf{q}), \quad (1.12)$$

$$\Delta G_\mathbf{k}(\mathbf{q}) \equiv G_{\mathbf{k}+\mathbf{q}/2}^R - G_{\mathbf{k}-\mathbf{q}/2}^A, \quad \Delta\Sigma_\mathbf{k}(\mathbf{q}) \equiv \Sigma_{\mathbf{k}+\mathbf{q}/2}^R - \Sigma_{\mathbf{k}-\mathbf{q}/2}^A, \quad (1.13)$$

which links the self-energy $\Sigma_\mathbf{k}$ and the irreducible vertex $U_{\mathbf{k}\mathbf{k}'}(\mathbf{q})$. The proof of Ward's identity in Ref. [21] implies that this identity is satisfied sequentially by going from diagram to diagram, thus calling for rigorous correspondence of the diagrams used in the calculation of $\Sigma_\mathbf{k}$ and $U_{\mathbf{k}\mathbf{k}'}(\mathbf{q})$. For example, the use of the first diagram in Fig. 1b requires using the first diagram in Fig. 2c; the use of the second diagram in Fig. 1b requires using the second, third, and fourth diagrams in Fig. 2c, and so on. Because of this, any approximation used for calculating the conductivity will not be self-consistent unless the corresponding approximation for the density of states has been formulated.

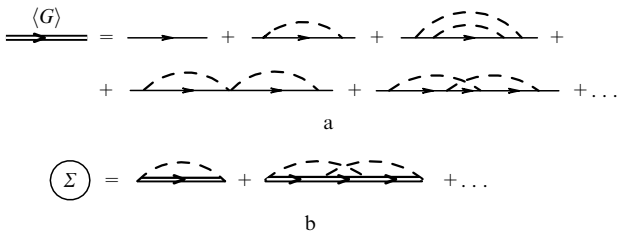

**Figure 1.** Diagrams for the mean Green function $\langle G \rangle$ (a) corresponding to the Gaussian random potential [9] or the Born approximation for randomly distributed impurities [14]; the series for the self-energy $\Sigma$ (b) only includes the irreducible diagrams.

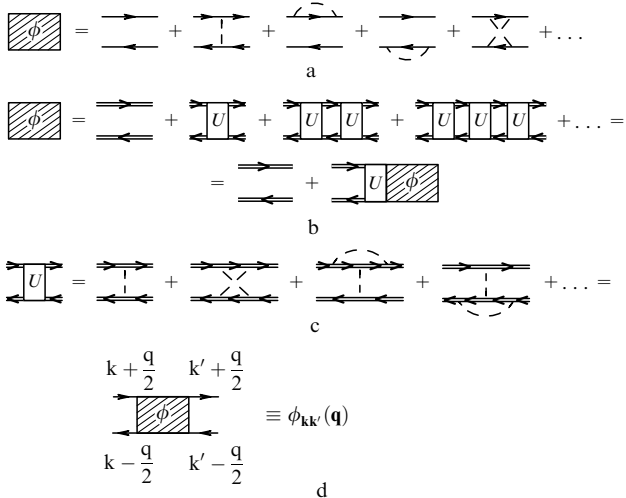

**Figure 2.** Diagram series for $\phi$ (a); graphic representation of the Bethe–Salpeter equation (b); definition of the irreducible vertex $U$ (c). Explanation of three-momentum notation (d).

The 1970s saw a gradual recognition [3, 22–24] of the fundamental nature of the problem of the Anderson transition and its profound linkage with the fluctuation theory of phase transitions [9–11]. This was most vividly symbolized by the discovery of the formal mathematical equivalence between the problem of calculating the mean Green's function for Eqn (1.4) with a Gaussian random potential (1.2) in the continuum limit $a_0 \to 0$, $a_0^d W^2 \to$ const, and the problem of the second-order phase transition with an $n$-component parameter of order $\boldsymbol{\varphi} = (\varphi_1, \varphi_2, \ldots, \varphi_n)$ in the limit $n \to 0$. This can be proved by comparing the diagram expansions (see p. 225 in Ref. [9]), or by going over to the functional integral using the replica method [25]. Then the coefficients in the Ginzburg–Landau integral

$$H\{\varphi\} = \int d^d x \left( \frac{1}{2} c|\nabla\boldsymbol{\varphi}|^2 + \frac{1}{2}\varkappa_0^2|\boldsymbol{\varphi}|^2 + \frac{1}{4}g_0|\boldsymbol{\varphi}|^4 \right) \quad (1.14)$$

are linked with the parameters of the disordered system by the relations

$$c = \frac{1}{2m}, \quad \varkappa_0^2 = -E, \quad g_0 = -\frac{W^2 a_0^d}{2}. \quad (1.15)$$

As usual, the coefficient at $|\nabla\boldsymbol{\varphi}|^2$ is positive, the coefficient at $|\boldsymbol{\varphi}|^2$ changes its sign in the neighborhood of the phase transition (which, given that the disorder is low and $d > 2$, occurs near the starting edge of the spectrum from where the energy $E$ is counted); the coefficient at $|\boldsymbol{\varphi}|^4$, however, has the 'wrong' sign. The latter circumstance obstructs a straightforward translation of the results of the theory of phase transitions to the physics of disordered systems: the traditional mean field theory is useless, and a consistent fluctuation treatment over the entire range of parameters is necessary†.

---

† When the diagrammatic technique is used, the negativity of $g_0$ is not important, since the expansion is performed in integer powers of $g_0$. The functional integrals with $g_0 < 0$ are interpreted as analytical continuation from positive $g_0$. As a matter of fact, as explained in Ref. [25], the use of the replica methods allows one to avoid the diverging functional integrals by virtue of an appropriate selection of the field $\varphi$; this points to the method of analytical continuation.

For calculating the correlator $\langle G^{\mathrm{R}} G^{\mathrm{A}} \rangle$ that defines the conductivity, one must use the effective Hamiltonian for two zero-component fields $\varphi$ and $\phi$ [25], which is also analyzed in the theory of phase transitions [26]:

$$H\{\varphi,\phi\} = \int \mathrm{d}^d x \left[ \frac{1}{2} c |\nabla \varphi|^2 + \frac{1}{2} c |\nabla \phi|^2 + \frac{1}{2} \varkappa_1^2 |\varphi|^2 \right.$$
$$\left. + \frac{1}{2} \varkappa_2^2 |\phi|^2 + \frac{1}{4} g_0 \left( |\varphi|^2 + |\phi|^2 \right)^2 \right], \quad (1.16)$$

where $\varkappa_1^2 = -E - \omega - \mathrm{i}\delta$, $\varkappa_2^2 = -E + \mathrm{i}\delta$.

By analogy with the theory of phase transitions, one can look forward to constructing a simple theory in a space of dimension $d = 4 - \epsilon$ for the Hamiltonians (1.14), (1.16). Such attempts were made in the mid-1970s, but failed to bear the desired results: the renormalization group analysis revealed the absence of fixed points lying within the range of applicability of the theory [25], and the parquet approach ran into the problem of the spurious pole [20, 5, 27]. Wegner's inequalities [28] for critical exponents pointed to the inadequacy of straightforward qualitative analogies with the theory of phase transitions.

Unavailing attempts at constructing a $(4 - \epsilon)$ theory [25, 27] promoted the opinion that the upper critical dimension $d_{\mathrm{c}2}$ for the Anderson transition (that is, the dimension above which the theory becomes much simpler) is other than four. Thouless [29] argued that the evidence pointing to the special role of the dimension $d = 4$ is genetically linked to the limiting transition to a 'Gaussian white noise' potential [that is, to the continuum limit $a_0 \to 0$, $a_0^d W^2 \to$ const in Eqn (1.4)], and only indicates that such a passage is nonphysical. His reasons drove some authors to the conclusions that $d_{\mathrm{c}2} = 6$ [30], 8 [31, 32], or even $\infty$ [33–35]. The lack of agreement on the value of $d_{\mathrm{c}2}$ went hand in hand with the lack of any constructive ideas on exactly how the condition $d > d_{\mathrm{c}2}$ may simplify the problem — that is, on the nature of the relevant 'mean-field theory'.

For the problem of the density of states near the Anderson transition, the issue of the upper critical dimension and the feasibility of $\epsilon$ expansion in its neighborhood has been exhaustively resolved in the author's recent papers [36–39]. We hold that $d_{\mathrm{c}2} = 4$, show how the condition $d > 4$ simplifies the problem, analyze the nature of the singularity at $d = 4$, and construct the $(4 - \epsilon)$-dimensional theory. This review is devoted to a systematization of these results. We pay more attention to the background of the problem (Sections 2, 3), describe the main ideas and the resulting qualitative picture (Section 4). In Section 5 we describe the methods for analyzing the higher-order perturbation theory, which can be used for a broad range of problems but currently seem to be underemployed. The structure of the $(4 - \epsilon)$-dimensional theory is described in Section 6; whenever possible, this is done without referring to the results for higher dimensions. Finally, in Section 7 we discuss the possible application of the $\epsilon$ expansion in the study of kinetic properties near the Anderson transition.

## 2. Special role of the dimension $d = 4$

In the case of weak disorder, the mobility edge lies in the neighborhood of the 'bare' edge of the spectrum, where the random potential may be regarded as Gaussian owing to the feasibility of averaging over the scales which are small compared to the electron wavelength but large with respect to the distance between the scatterers (the so-called Gaussian portion of the spectrum [40]). In high-dimensional spaces, the discrete nature of the lattice is of primary importance, so further on we shall use the Anderson model (1.1) with a Gaussian distribution for the energy of the sites (1.2); the disorder is assumed to be small, and our concern is with the low energy range

$$W \ll J, \quad |E| \ll J, \quad (2.1)$$

where $E$ is counted from the lower edge of the band.

The special role of the dimension $d = 4$ for the Gaussian model (1.1) is manifested in a number of aspects.

**Ioffe–Regel criterion.** The absence of localization is expressed by the known Ioffe–Regel criterion [2]

$$p l \gtrsim 1 \quad \text{or} \quad E\tau \gtrsim 1, \quad (2.2)$$

where $E$ and $p$ are the energy and momentum of an electron, and $l$ and $\tau$ are the mean free path and time. In the Born approximation,

$$\tau^{-1}(E) \sim a_0^d W^2 \nu_0(E), \quad (2.3)$$

where

$$\nu_0(E) \sim (a_0^d J)^{-1} \left(\frac{E}{J}\right)^{(d-2)/2} \quad (2.4)$$

is the density of states of an ideal lattice, and at $d < 4$ the condition (2.2) reduces to

$$E \gtrsim J \left(\frac{W}{J}\right)^{4/(4-d)}, \quad (2.5)$$

whereas at $d > 4$ this condition under our assumptions holds for all $E$.

**Analogy with the theory of phase transitions.** The problem of the Anderson transition can be reformulated in a formally exact way in terms of effective Hamiltonians (1.14), (1.16), which only differ in the sign of $g_0$ from the effective Hamiltonians in the theory of second-order phase transitions, for which the role of the dimensionality $d = 4$ is well known: for $d > 4$, the Landau mean-field theory becomes exact. For $g_0 < 0$, the mean field theory makes no sense, although some indications of the retention of the special role of $d = 4$ can be derived from the Wilson renormalization group (Fig. 3). For all $d$, the renormalization group transformation exhibits a Gaussian fixed point $g_0 = 0$; in addition, for $d$ close to 4, one can prove the existence of a nontrivial fixed point $g_0 = g^*(d)$. At $d = 4$, these two fixed points 'exchange instability': the former is stable and the latter unstable at $d > 4$, and at $d < 4$, the situation is reversed. The arrows in Fig. 3 indicate the change in the 'charge' $g_0$ in the renormalization group transformations. If the initial value of $g_0$ is negative, then at $d < 4$ the Wilson fixed point $g^*$ is unattainable, which manifests the inconsistency of straightforward analogies with the theory of phase transitions [25]. However, with $d > 4$ and small negative $g_0$ the system approaches the Gaussian fixed point, which means that the theory is relatively simple.

In reality, the situation is more complicated: for negative $g_0$ one must include the nonperturbative terms into the equations of the Wilson renormalization group, which lead to a complete destruction of the latter. Nevertheless, the arguments developed above clearly demonstrate the difference in the transformation properties for dimensions greater and less than four.

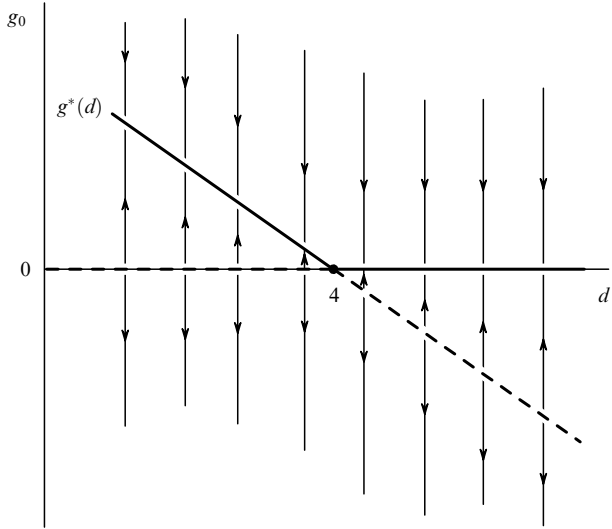

**Figure 3.** Change of the 'charge' $g_0$ caused by the renormalization group transformations.

**Method of optimal fluctuation.** For large negative $E$, the properties of localized states can be studied with Lifshitz's method of optimal fluctuation [8, 40]. By virtue of Eqn (1.2), the probability of fluctuation occurrence for a potential well of depth $V$ and radius $R$ is of the order† of

$$P(V, R) \sim \exp\left(-\frac{V^2 R^d}{W^2 a_0^d}\right). \tag{2.6}$$

If the well exhibits a level $E = -|E|$, the parameters $V$ and $R$ are linked by

$$E = -V + \frac{1}{mR^2} \approx -V + J\left(\frac{a_0}{R}\right)^2, \tag{2.7}$$

which allows the elimination of $V$ from Eqn (2.6):

$$P(E, R) \sim \exp\left\{-\left(\frac{R}{a_0}\right)^d \left[\frac{|E| + J(a_0/R)^2}{W}\right]^2\right\}$$
$$\equiv \exp\left[-S(E, R)\right]. \tag{2.8}$$

The total probability $P(E)$ of the occurrence of level $E$, which determines the density of states $\nu(E)$, is found by integrating Eqn (2.8) with respect to $R$, which in the saddle-point approximation reduces to the replacement of $R$ with $R_0$ — the point of minimum of $S(E, R)$. At $d < 4$, the radius of the optimal fluctuation is $R_0 \sim |E|^{-1/2}$ (Fig. 4a) and diverges when $|E| \to 0$, which leads to the well-known Lifshitz law

$$\nu(E) \sim \exp\left[-\text{const}\,|E|^{(4-d)/2}\right]. \tag{2.9}$$

At $d > 4$ (Fig. 4a), the extremum of $S(E, R)$ corresponds to the smallest possible $R$, that is, $R_0 \sim a_0$, whence

$$\nu(E) \sim \exp\left[-\frac{(J + |E|)^2}{W^2}\right]. \tag{2.10}$$

† All coefficients of the order of unity are dropped in the estimates to follow.

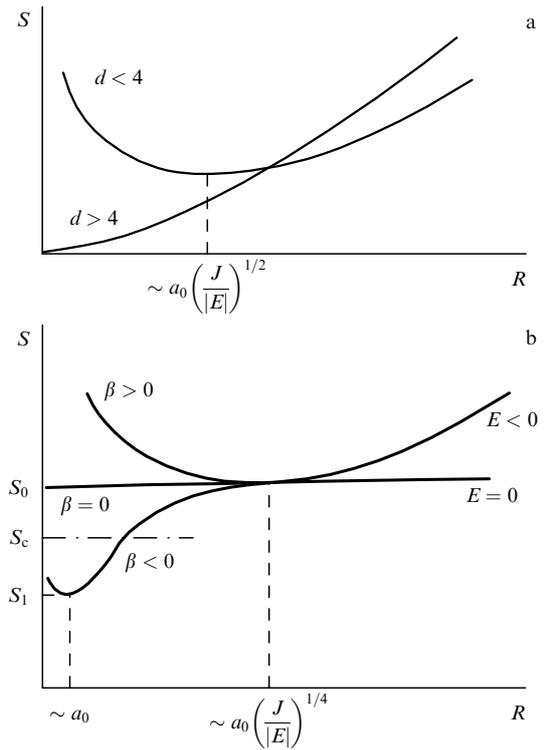

**Figure 4.** Function $S(E, R)$ vs. $R$ with $E = \text{const}$: (a) $d > 4$ and $d < 4$; (b) $d = 4$.

Since the extremum is reached at the boundary of the domain of definition, the derivative of $S(E, R)$ with respect to $R$ does not vanish. In the field-theory formulation [20, 41–43], this corresponds to the absence of classical solutions — the instantons [44].

At $d = 4$ (Fig. 4b), we have $S(E, R) = \text{const} = S_0$ at $E = 0$, and the situation is close to degeneracy. At large $R$, the degeneracy is removed on account of $E$ being finite, $S(E, R) - S_0 \sim E^2 R^d$; at small $R$, the behavior of the spectrum $\epsilon(p)$ at large values of $p$ is important. If the expansion $\epsilon(p)$ in $p$ involves terms that are quartic in $p$, in addition to terms quadratic in $p$, then in place of Eqn (2.7) we obtain

$$E = -V + J\left(\frac{a_0}{R}\right)^2 + \beta J\left(\frac{a_0}{R}\right)^4. \tag{2.11}$$

At $\beta > 0$, the function $S(E, R)$ deviates from $S_0$ upwards, which ensures a minimum at $R_0 \sim |E|^{-1/4}$. At $\beta < 0$, the deviation is downwards, and the minimum is attained at $R_0 \sim a_0$, where the higher terms of the expansion $\epsilon(p)$ in $p$ gain importance (Fig. 4b). We see that the transition from higher to lower space dimensionalities 'continues' at $d = 4$ with respect to the parameters of the model: for $\beta < 0$, the optimal fluctuation is localized on the atomic scale, similarly to the case of $d > 4$, whereas for $\beta > 0$, the radius of optimal fluctuation diverges at $|E| \to 0$, which is characteristic of lower dimensionalities. Accordingly, the asymptotics of the fluctuation tail at $E \to -\infty$ are different:

$$\nu(E) \sim \begin{cases} \exp\left[-\dfrac{J^2}{W^2}\left(1 + \dfrac{|E|}{J}\right)\right], & \beta < 0, \\ \exp\left[-\dfrac{J^2}{W^2}\left(1 + \dfrac{|E|^{1/2}}{J^{1/2}}\right)\right], & \beta > 0. \end{cases} \tag{2.12}$$

In the range of small $E$ that we are interested in, where the mobility edge occurs, the boundary between the two models is no longer distinct. As a matter of fact, at $\beta < 0$ there is competition between the contributions from the minimum with $S = S_1$ and the overlying plateau $S(E, R) = S_0$, whose width grows indefinitely as $|E|$ decreases. Integration of $P(E, R)$ with due account for both contributions results in

$$P(E) \sim \nu(E) \sim \exp(-S_1) + \left(\frac{J}{|E|}\right)^\alpha \exp(-S_0), \qquad (2.13)$$

where $\alpha = 1/2$. As $S_1$ increases, the second term (the plateau) starts to dominate sooner than $S_1$ becomes equal to $S_0$, and the transition to $\beta > 0$, accompanied by the disappearance of the first term in Eqn (2.13), is of little consequence. The above value of the exponent $\alpha$ cannot be taken seriously, since the accuracy of the method does not allow for evaluating the pre-exponential factor. Its exact value can be derived from considerations of renormalizability and is equal to $1/3$ (see below).

The damping $\Gamma$, determined by the imaginary part of the self-energy $\Sigma(p, \varkappa)$ at $p = 0$ (where $\varkappa$ is the renormalized value of $\varkappa_0$), in the domain of applicability of the method of optimal fluctuation is proportional to the density of states $\nu(E)$, and with due account for the dimensionality is given by

$$\Gamma \sim J\left[\exp(-S_1) + \left(\frac{J}{|E|}\right)^{1/3} \exp(-S_0)\right]. \qquad (2.14)$$

The energy always occurs in the combination $E + i\Gamma$, and in the neighborhood of the Anderson transition we may replace $|E|$ with $\Gamma$; it is easy to see that the first term in braces dominates at $S_1 < 3S_0/4$, and the second dominates when the reverse is true. Since $S(E, R) \sim W^{-2}$ [see Eqn (2.8)], in the limit of weak disorder there is a sharp distinction between the two types of models: at $S_1 < S_c$, the optimal fluctuation is determined by the atomic scale, and the discreteness of the lattice is of fundamental importance, as in the case of $d > 4$; at $S_1 > S_c$, the large-radius fluctuations are definitive, and the treatment can be based on the continuum model with a square-law spectrum: the situation is similar to that for lower dimensionalities.

**Renormalizability of the theory.** The above classification of models is directly related to the renormalizability of the theory. The diagram of $N$th order for the self-energy $\Sigma$ has a dimensionality of $p^r$ with respect to momentum, where $r = 2 + (d - 4)N$. At $d > 4$, the degree of divergence at large momenta increases with the order of the diagram, and the theory is not renormalizable [45]: one must explicitly define the cutoff parameter $\Lambda$, which points to the importance of the structure of the Hamiltonian on the atomic scale. At $d < 4$, we have $r < 2$ for all $N$: if we subtract from each diagram its value at $p = \varkappa = 0$, the exponent $r$ is reduced by two, and the difference $\Sigma(p, \varkappa) - \Sigma(0, 0)$ does not contain divergences, which are absorbed by the quantity $\Sigma(0, 0)$, which only shifts the origin of the energy. At $d = 4$, the difference $\Sigma(p, \varkappa) - \Sigma(0, 0)$ contains logarithmic divergences, which can be removed by renormalizing the charge and the Green's function [45, 46]; one must bear in mind, however, that the standard proofs of renormalizability only deal with the range of distances greater than $\Lambda^{-1}$. By assumption, the scales below $\Lambda^{-1}$ do not give $\delta$-shaped contributions which are significant at $\Lambda \to \infty$. Our estimates show that this is not always the case: the renormalizable long-range contribution (the plateau) only dominates when $S_1 > S_c$; otherwise it is small compared with the nonrenormalizable short-range contribution.

In this way, there are four fundamentally different classes of theories:

(a) nonrenormalizable theories at $d > 4$;

(b) nonrenormalizable theories in the logarithmic situation ($d = 4$, $S_1 < S_c$);

(c) renormalizable theories in the logarithmic situation ($d = 4$, $S_1 > S_c$);

(d) theories renormalizable by means of one subtraction (super-renormalizable theories) at $d < 4$.

Simplifications arising in the higher dimensionalities can be used for the construction of the $(4 - \epsilon)$-dimensional theory but this requires a consistent treatment of all four types of theories: the passage from higher to lower dimensionalities is described in Section 4.

Of course, the special role of the dimensionality $d = 4$ did not pass unnoticed, but was often interpreted in the wrong way. The apparent fulfillment of the Ioffe – Regel condition for all positive $E$ and the absence of instantons in the continuum limit [44] has led to the belief, common in the early 1970s, that there are no localized states at $d = 4$ at all (Ref. [2], p. 22). The paper by Thouless [29] was mainly concerned with the critical revision of this viewpoint: a more careful analysis reveals that the localized states do exist, but the passage to the Gaussian limit of white noise $a_0 \to 0$, $a_0^d W^2 \to \mathrm{const}$ smears them over the semiaxis of negative energies with zero density, which leads to the singularity of $\nu(E)$ at the mobility edge. Any model that has a minimum length scale restores the regularity of the density of states, which brought Thouless to the conclusion (although very cautiously worded†) that the space dimensionality of $d = 4$ is not the upper critical value. The weakness of this standpoint is clear from the fact that all the above features manifest even before the transition to the white noise limit.

## 3. Problem of the spurious pole

Attempts to construct a $(4 - \epsilon)$ expansion for the Anderson transition were made in the mid-1970s; however, they ran into serious difficulties which turned out to be profoundly linked with other problems of theoretical physics. We know that in quantum electrodynamics there is a relation which links the observed charge $e$ that enters into the Coulomb law with the 'bare' charge $e_0$ that occurs in the initial Lagrangian [45, 47]

$$e^2 = \frac{e_0^2}{1 + (2e_0^2/3\pi)\ln(\Lambda/m)}, \qquad (3.1)$$

where $m$ is the mass of electron. The cutoff parameter $\Lambda$ has no physical meaning and must be directed to infinity, then the observed charge $e \to 0$ at any value of the bare charge $e_0$. This so-called zero-charge situation was at first regarded as a paradox evincing some fundamental flaws in the theory [48]. According to the modern view, $e_0$ must be interpreted as the effective charge related to the scale $\Lambda^{-1}$ and depending on $\Lambda$ in such a way that the left-hand side of Eqn (3.1) always

---

† "... Both the density of states and the properties of localized states are special in four dimensions ..., but these special features do not seem to make the problems we are interested in any simpler, and so extrapolation from four to three dimensions is not obviously helpful" ([29], p. L603).

represents the observed charge $e$, that is,

$$e_0^2(\Lambda) = \frac{e^2}{1-(2e^2/3\pi)\ln(\Lambda/m)}.  \quad (3.2)$$

As a result, the charge of the electron starts to increase when the length scale decreases below the Compton length $m^{-1}$ ($\equiv \hbar/mc$) (Fig. 5). In quantum chromodynamics, the formula similar to Eqn (3.1) has a plus sign in the denominator, and the effective charge increases instead of decreasing as the length scale increases (see Fig. 5). This explains why quarks behave as free particles at small distances (*asymptotic freedom*), whereas at large distances the interaction is so strong as to prevent their escapement (*confinement*). The effective charge as a function of distance can be described on the basis of Eqn (3.1) only provided it is small enough. What expression replaces Eqn (3.1) when the charge becomes of the order of 1 [which corresponds to small distances in quantum electrodynamics, and large in quantum chromodynamics (see Fig. 5)] is still not known [49].

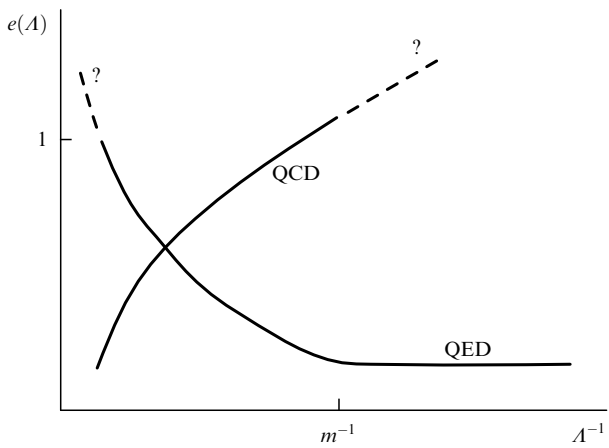

**Figure 5.** Effective charge as a function of length scale in quantum electrodynamics (QED) and quantum chromodynamics (QCD).

The linkage between the problem of Anderson transition and the theory of interacting particles relates to the fact that in quantum field theory the model (1.14) with $d=4$, $n=1$ corresponds to the well-known $\varphi^4$ model of a relativistic Bose gas with point interaction [45, 50]; the quantity $\varkappa$ represents the mass of the particle, and the negative values of the interaction constant $g_0$ (attraction) correspond to the unstable field theory. The situation with renormalizability in the $\varphi^4$ model is the same as that in quantum electrodynamics, and a relation similar to Eqn (3.1) links the renormalized charge $g$, which describes the interaction at large distances, with its bare value $g_0$ [11, 45, 51, 52]:

$$g = \frac{g_0}{1+K_4(n+8)g_0\ln(\Lambda/\varkappa)},  \quad (3.3)$$

where $K_4 = (8\pi^2)^{-1}$ is the area of a unit four-dimensional sphere divided by $(2\pi)^4$. The result (3.3) is obtained by summation of the so-called parquet diagrams which involve the highest power of the major logarithm [52–55]. At $g_0 > 0$ — that is, in the conventional theory of phase transitions — the effective interaction described by Eqn (3.3) tends to zero as $\varkappa \to 0$, upon approaching the point of transition. We are dealing then with the zero-charge situation, which in this case takes place literary, since the bare charge $g_0$ and the cutoff parameter $\Lambda$ correspond to the atomic scale and are observable quantities. Upon transition from $d=4$ to $d=4-\epsilon$, the charge $g$ in the limit $\varkappa \to 0$ assumes a small but finite value. As a matter of fact, it is the weakness of the effective interaction that ensures the success of Wilson's $\epsilon$-expansion, the results of which are perfectly well reproduced in the parquet formulation of the theory [52, 56].

The use of the parquet approximation in the theory of disordered systems leads to an expression similar to Eqn (3.3) with $g_0 < 0$ [5, 20, 27], and the situation is asymptotically free: upon approaching the mobility edge, the effective interaction is growing rather than decreasing, and at a certain small $\varkappa$ (that is, a small value of the renormalized energy $E$), expression (3.3) exhibits a spurious pole, which cannot be eliminated in the context of the parquet approximation. Formal use of the parquet results, like Eqn (3.3), leads to the divergence of physical quantities. Thus, we are dealing with a paradoxical situation: one and the same approximation in two mathematically equivalent problems leads to a practically complete solution of the problem in one case (the theory of phase transitions), and to obviously nonphysical results in the other (the theory of disordered systems). According to Sadovsky [5, 20], the problem of the spurious pole is the main obstacle in the construction of a consistent theory of Anderson transition. Here, we are going to present a comprehensive solution to this problem as applied to the calculation of the density of states.

## 4. From higher to lower dimensions

### 4.1. Simplification of theory at $d > 4$

From the diagram series for self-energy $\Sigma(p, E)$ (Fig. 1b), it is easy to see that the increase in the order of the diagram by one (the order of diagram is defined by the power of $g_0$, that is, by the number of impurity lines) gives rise to an extra factor of $W^2$ and two additional $G$ functions. In rough estimations, one may use for the latter the functional form obtained in the domain of weak disorder† [14]

$$G(p, E) = \frac{1}{E - p^2/2m + \mathrm{i}\Gamma}.  \quad (4.1)$$

For $d < 4$, the divergence at large momenta is eliminated in view of the renormalizability of the theory, and the integrals are determined by the small values of $p$. Because of this, in the range of small $E$ that we are concerned with, the Green's function ought to be regarded to be of the order of $1/\Gamma$. In the case of weak disorder, the damping $\Gamma$ is small, being determined by parameter $W$, and, occurring in the denominator, may compensate for the smallness of $W^2$. A detailed analysis reveals that this is what actually happens in the neighborhood of the mobility edge, so that with $|E| \sim \Gamma$ the actual parameter of expansion is of the order of unity.

With $d > 4$, the situation is changed. Since the theory is nonrenormalizable, the integrals are defined by the large momenta $p \sim \Lambda$, and the Green's function (4.1) is of the order of $1/J$. Because of this, the parameter of expansion in the domain of weak disorder is small,

$$\frac{W^2}{J^2} \ll 1.  \quad (4.2)$$

---

† Wherever relevant, it is the retarded Green's function.

The existence of the small parameter does not, however, immediately simplify the problem, because the attempt to restrict oneself to a finite-order perturbation theory leads to the loss of the fluctuation tail. An approximation that gives asymptotically exact results over the entire range of energies was constructed by the author in Ref. [36]: the perturbation theory expansion for $\Sigma$ must be approximated by the first term and the sum of remote terms, from some large $N_0$ to infinity. The high-order terms give the contribution which is associated with the divergence of the series and therefore does not depend on $N_0$.

The divergence of the perturbation theory series is directly related to the existence of the fluctuation tail of the density of states. Indeed, for a Gaussian distribution of the energy of lattice sites, there is a finite probability of existence of arbitrarily deep fluctuations of potential, and therefore arbitrarily deep energy levels for an arbitrarily small value of $W$. In other words, the density of states $v(E) \sim \operatorname{Im} G(E + \mathrm{i}\delta)$ is nonzero for all $E$ and $W$; hence

$$G(E + \mathrm{i}\delta) - G(E - \mathrm{i}\delta) = \mathrm{const} \times v(E) \neq 0$$

$$\text{for all} \quad E, \ g_0 < 0, \qquad (4.3)$$

and the exact Green's function $G(E)$ has the discontinuity at negative $E$, which did not occur in the unperturbed Green's function $G_0(E)$. According to the well-known theorem of calculus, the sum of a series composed of continuous functions is continuous if the series is uniformly convergent; the series is uniformly convergent if it is majorizable by a convergent numerical series [57]. If the coefficients of expansion of $G(E)$ in powers of $g_0$ grow slower than $a^N$ with a certain finite $a$, then for small $|g_0|$ the series is majorizable by a convergent geometric progression and is uniformly convergent. Then, the continuity of terms of the series, which follows from the continuity of $G_0(E)$, implies that Eqn (4.3) does not hold. Therefore the coefficients of the expansion grow faster than $a^N$ for arbitrarily large $a$, and the sequence diverges for arbitrarily small $|g_0|$. As a matter of fact, the divergence is factorial, because the number of diagrams of the same order is factorially large.

The remote terms of the perturbation theory series can be calculated using the Lipatov method [58], which is based on the following simple idea. The coefficients $F_N$ of expansion of the function $F(g)$

$$F(g) = \sum_{N=0}^{\infty} F_N g^N \qquad (4.4)$$

may be represented as

$$F_N = \int_C \frac{\mathrm{d}g}{2\pi\mathrm{i}} \frac{F(g)}{g^{N+1}}, \qquad (4.5)$$

where the contour $C$ encloses the point $g = 0$ on the complex plane. Rewriting the denominator as $\exp[-(N+1)\ln g]$, for large $N$, we get an exponential with a large exponent, which indicates that the saddle-point method might work. Now, we apply Eqn (4.5) to the functional integral

$$F(g) = \int \mathrm{D}\varphi \exp(-H\{g, \varphi\}),$$
$$H\{\varphi\} = H_0\{\varphi\} + gH_{\mathrm{int}}\{\varphi\}. \qquad (4.6)$$

Then

$$F_N = \int_C \frac{\mathrm{d}g}{2\pi\mathrm{i}} \int \mathrm{D}\varphi \exp\bigl[-H_0\{\varphi\} - gH_{\mathrm{int}}\{\varphi\} - (N+1)\ln g\bigr]. \qquad (4.7)$$

The idea of the Lipatov method is that the saddle-point in Eqn (4.7) is sought with respect to $g$ and $\varphi$ *simultaneously*; it exists for most Hamiltonians of practical interest $H\{g, \varphi\}$, and is realized on a certain space-localized function $\varphi(x)$ (known as an *instanton*); as it happens, the conditions of applicability of the saddle-point method are satisfied for large $N$ irrespective of its workability for the initial integral (4.6). This completely changes the situation: while the exact calculation of functional integrals is, as a rule, not possible, they can always be calculated in the saddle-point approximation.

The diagram series for the Green's function of a disordered system (Fig. 1a) is represented with the aid of the replica method as a functional integral with Hamiltonian (1.14), which allows us to calculate its remote terms using the Lipatov approach. The transition from the expansion for $G(p, E)$ to the expansion for the self-energy $\Sigma(p, E)$ does not pose any problems, since the straightforward algebra developed for the factorial series allows them to be manipulated as easily as if they were finite expressions (see Section 5.3). For the $N$th coefficient in the expansion $\Sigma(p, E)$, we get the functional form

$$c\Gamma(N+b)a^N, \qquad (4.8)$$

where $\Gamma(x)$ is the gamma function, $a = a(E)$, $b = \mathrm{const}$, and $c = c(p, E)$. The divergent series is formally summed by representing the gamma function as a definitive integral and finding the sum of the resulting geometric progression†:

$$\sum_{N=N_0}^{\infty} c\Gamma(N+b)(ag_0 + \mathrm{i}\delta)^N$$

$$= \sum_{N=N_0}^{\infty} c \int_0^{\infty} \mathrm{d}x \, \exp(-x) x^{N+b+1}(ag_0 + \mathrm{i}\delta)^N$$

$$= c \int_0^{\infty} \mathrm{d}x \, \exp(-x) x^{b-1} \frac{(xag_0)^{N_0}}{1 - x(ag_0 + \mathrm{i}\delta)}$$

$$= c \int_0^{\infty} \mathrm{d}y \, \frac{\exp(-1/y)}{y^b} \left(\frac{ag_0}{y}\right)^{N_0} \frac{1}{y - ag_0 - \mathrm{i}\delta}. \qquad (4.9)$$

The infinitesimal imaginary term $\mathrm{i}\delta$ in Eqn (4.9) comes from two sources (with the same sign from both of these): from the imaginary term $+\mathrm{i}\delta$ or $-\mathrm{i}\delta$, which must be added to the energy $E$ so as to particularize the selection of the Green's function (retarded or advanced), and from the imaginary addition to $g_0$, which specifies the method of analytical continuation from positive to negative values of $g_0$, which is performed via the upper or lower half-plane for, respectively, the retarded or advanced Green's function [25]. It is easy to check that the real part of sum (4.9) with small $g_0$ is well approximated by the first term in the series, and it is actually not necessary to add up the remote terms. A more interesting result is obtained for the imaginary part of the sum:

$$\mathrm{Im} \sum_{N=N_0}^{\infty} c\Gamma(N+b)(ag_0 + \mathrm{i}\delta)^N = \frac{\pi c}{(ag_0)^b} \exp\left(-\frac{1}{ag_0}\right),$$

$$ag_0 > 0. \quad (4.10)$$

---

† It is well known that if any formal manipulations with divergent series are permitted, the sum can be reduced to any desired value. The theory of divergent series [59] is based on the idea that certain restrictions must be imposed on such manipulations, so as to make the sum of the series independent of the method of summation. A complete solution of this problem (in the form of necessary and sufficient conditions) has apparently not been found as yet, although practical methods are numerous enough: there is a long list of 'good' transformations for which the equivalence theorems have been proved, so that if a divergent series has been reduced to a finite sum through manipulations from this list, the result will not depend on the method of summation. As a matter of fact, in Eqn (4.5) we use the well-known Borel transformation, which is included in that list.

Expression (4.10) displays several peculiar features:

(a) Each term on the left-hand side of Eqn (4.10) has an infinitesimal imaginary part $\sim \mathrm{i}\delta$, which together give a finite contribution to the sum of the series.

(b) The right-hand side of Eqn (4.10) contains a typical 'nonperturbative' contribution which cannot be expanded in the conventional power series, but, as we readily see, is perfectly well represented by a divergent series. As a matter of fact, Eqn (4.10) provides a mechanism for extraction of nonperturbative contributions from the diagram technique, the feasibility of which was challenged even in very creditable publications (see, for example, Ref. [60]).

(c) The right-hand side of Eqn (4.10) does not depend on $N_0$; in other words, the nonperturbative contribution comes from the range of arbitrarily large $N$.

(d) As we shall see, the coefficient $a$ is negative for Hamiltonian (1.14), and the nonperturbative contribution only arises when $g_0 < 0$. This explains why in the conventional theory of phase transitions a disregard of the factorial divergence of the perturbation theory series does not entail any grave consequences.

The inclusion of the nonperturbative contribution restores the fluctuation tail of the density of states which was lost when the series was truncated at a finite number of terms: even from Eqn (4.10) we see that the functional form of the right-hand side complies with the results (2.10), (2.12) of the method of optimal fluctuation — this reconfirms on the quantitative level the above-mentioned linkage between the divergence of the series and the existence of the fluctuation tail. Later on, this linkage will be demonstrated at an even higher level: the instanton equation in the Lipatov method will coincide with the equation of the typical wave function in the field of optimal fluctuation (see Chapter 4 in Ref. [40]). Accordingly, the above classification of models (Fig. 4a, b) will be manifested in yet another fundamental aspect, the nature of the divergence of the perturbation series. The existence of a nonperturbative contribution removes all the inconsistencies noted above. As a result, it becomes possible to calculate the density of states at all energies and to prove its regularity at the mobility edge.

### 4.2 Four-dimensional nonrenormalizable (lattice) models

At $d = 4$, all diagrams for the self-energy $\Sigma(p, \varkappa)$ (where $\varkappa$ is the renormalized value of $\varkappa_0$) diverge quadratically. Quadratic divergences are eliminated by subtracting from each diagram its value at $p = 0$, $\varkappa = 0$; the difference $\Sigma(0, \varkappa) - \Sigma(0, 0)$ diverges only logarithmically. Arranging the contributions of the diagrams in the powers of logarithms, for $p = 0$ we get

$$\Sigma(0, \varkappa) = \Sigma(0, 0) + \varkappa^2 \sum_{N=1}^{\infty} g_0^N \sum_{K=0}^{N} A_N^K \left(\ln \frac{\Lambda}{\varkappa}\right)^K,$$

$$\Sigma(0, 0) = \Lambda^2 \sum_{N=1}^{\infty} B_N g_0^N. \quad (4.11)$$

The $N$th-order contribution involves all powers of logarithms from zero to a certain highest order which is determined by the parquet graphs and is currently equal to the order of diagram $N$. In the $(N, K)$ plane, each integer-valued point that lies not higher than the main diagonal corresponds to the coefficient $A_N^K$ (Fig. 6a). The exact solution of the problem requires the knowledge of all these coefficients; our aim, however, consists in obtaining the results which would be asymptotically accurate in the limit of weak disorder. Such results are obtained in the theory of phase transitions ($g_0 > 0$) by retaining in Eqn (4.11) only the parquet coefficients $A_N^N$ corresponding to the highest powers of large logarithms; they are easily found using results of Ginzburg paper [37, 52]:

$$A_N^N = \left[-K_4(n+8)\right]^N \frac{\Gamma(N-\beta_0)}{\Gamma(N+1)\Gamma(-\beta_0)}$$

$$\xrightarrow{N\to\infty} \frac{1}{\Gamma(-\beta_0)} \left[-K_4(n+8)\right]^N N^{-\beta_0 - 1}, \quad (4.12)$$

where $\beta_0 = (n+2)/(n+8)$. Expression (4.12) reveals why the parquet approximation is inadequate with $g_0 < 0$: the parquet coefficients do not exhibit factorial growth, and if they alone are used, the fluctuation tail is lost. From the above proof of divergence of the series it follows that the terms with lower powers of logarithms grow faster and begin to dominate at large $N$.

In principle, it is possible to construct the second, third, etc. logarithmic approximations by including the coefficients $A_N^{N-1}$, $A_N^{N-2}$, etc. in Eqn (4.12). These coefficients are implicitly defined in the calculation of the subsequent terms of the $\epsilon$-expansion by Wilson's method based on the existence of the exact renormalization group at $d = 4 - \epsilon$, $g_0 > 0$ (see Chapter 9 in Ref. [9]). For finite $K$ and $N \to \infty$, we have

$$A_N^{N-K} = \mathrm{const}\left[-K_4(n+8)\right]^N N^{-\beta_0 - 1} (N \ln N)^K. \quad (4.13)$$

We see that factorial growth does not occur in any finite logarithmic approximation. It can only be surmised at $K \sim N$, when Eqn (4.13) is no longer valid.

Information concerning the fastest-growing coefficients can be obtained by Lipatov's method; the situation is quite different for renormalizable and nonrenormalizable classes of models. In the nonrenormalizable models, the predominant contribution comes from the minimum $S = S_1$ (Fig. 4b), and the discreteness of the lattice is of fundamental importance: Lipatov's asymptotics have to be calculated for the lattice version of Hamiltonian (1.14). As it turns out, the highest growth rate at $N \to \infty$ is displayed by the terms with the zeroth and first powers of logarithms, corresponding to the coefficients $A_N^0$, $A_N^1$, and $B_N$, whereas the terms with the higher powers of logarithms grow at a slower rate and are not represented by the principal asymptotics. The fact is that the contribution from the minimum $S = S_1$ at $|E| \sim \Gamma$ is practically independent of the energy [see the first expression in Eqn (2.12)], and ought to be mainly determined by the coefficients $B_N$ in expansion (4.11); the coefficients $A_N^0$ and $A_N^1$ give a weak dependence on energy, which only becomes significant at large negative values of $E$.

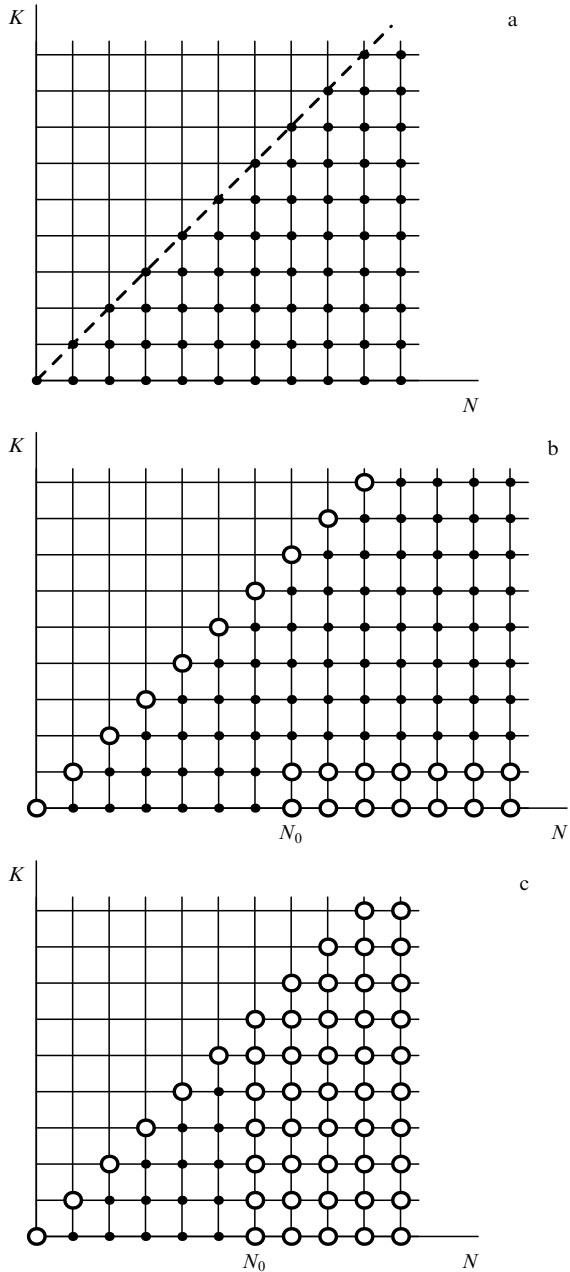

**Figure 6.** (a) In the $(N, K)$ plane, each integer-value point not above the main diagonal corresponds to a term $\sim g_0^N \ln^K(\Lambda/\varkappa)$ in the sum (4.11); (b) terms that must be taken into account (empty circles) to obtain asymptotically exact results (in the limit of weak disorder) for the nonrenormalizable class of models; (c) the same for the renormalizable models.

To obtain an asymptotically accurate description of the entire energy range in the limit of weak disorder, including the neighborhood of the Anderson transition, expansion (4.11) must include (Fig. 6b) (a) the parquet terms, defined by the coefficients $A_N^N$, as the terms containing the highest power of the large logarithm, and (b) starting with a certain large number $N_0$, the fastest-growing terms corresponding to the coefficients $A_N^0$, $A_N^1$, and $B_N$. The importance of the latter is related to the divergence of the series, and the selection of $N_0$ is unessential.

Then, from Eqn (4.11) we get

$$\Sigma(0, \varkappa) - \operatorname{Re} \Sigma(0, 0)$$
$$= \varkappa^2 \left[ \left( 1 + 8K_4 g_0 \ln \frac{\Lambda}{\varkappa} \right)^{1/4} - 1 \right] + \mathrm{i} \Gamma_0(\varkappa), \quad (4.14)$$

which only differs from the parquet approximation (the first term on the right-hand side) by the presence of an exponentially small nonperturbative term $\mathrm{i}\Gamma_0(\varkappa)$. Disregarding the weak dependence of the latter on $\varkappa$, we may interpret this term as representing the imaginary part of the quantity $\Sigma(0,0)$, which in the conventional theory of phase transitions ($g_0 > 0$) determines the location of the transition point. We see that the passage to negative values of $g_0$ results mainly in the occurrence of the imaginary part of the 'temperature' of transition. In other words, the physical variables of a disordered system, defined by the average Green's function, are described by the formulas of the theory of phase transitions with complex $T_c$. This circumstance ensures evasion of the spurious pole and regularity of the density of states at all energies.

### 4.3 Four-dimensional renormalizable models
The approximation used above fails when $S_1$ approaches $S_c$ [37, 38]:

(a) the definitive equation for $\Gamma(E)$ gives physically meaningless solutions at $S_1 > S_c$;

(b) at $S_1 \to S_c$, the contribution of subleading logarithmic terms defined by the coefficients $A_N^{N-K}$ with $K \sim 1$ sharply increases;

(c) at $S_1 \approx S_c$, the contribution from the plateau becomes important (Fig. 4b), whose strong dependence on energy points to the growing significance of the coefficients $A_N^K$ with $K \neq 0$.

Thus, while the lattice models are dominated by the most 'senior' and most 'junior' logarithms, the contributions of all $K$ are generally important in sum (4.11) upon transition to the renormalizable models.

In the latter case we come to the following statement of the problem. We select an integer number $N_0$, which is large compared to 1 but small compared to the large parameters of the theory. For $N < N_0$ we only retain in Eqn (4.11) the parquet coefficients $A_N^N$ distinguished by the large logarithms; at $N \geqslant N_0$, all terms are generally important in the sum over $K$ (Fig. 6c), but the condition $N \gg 1$ allows us to calculate the coefficients $A_N^K$ in the principal asymptotics with respect to $N$.

In renormalizable models, it is the plateau contribution that dominates (Fig. 4b); in other words, the large-radius instantons are significant, and the calculation of Lipatov's asymptotics can be carried out in the continuum model. The contribution of the $N$th order into $\Sigma(0, \varkappa)$ has the form

$$\varkappa^2 g_0^N c_2 \Gamma(N+b) a^N (\ln N)^{-\gamma} \exp\left( \frac{\sigma \ln \Lambda}{\varkappa} \right), \quad (4.15)$$

where $a$, $b$, $c_2$, and $\sigma$ are constants of order 1, defined in Section 5.5. Comparing this with expansion (4.11), we get

$$A_N^K = \frac{\sigma^K}{K!} A_N^0, \qquad A_N^0 = c_2 \Gamma(N+b) a^N (\ln N)^{-\gamma}. \quad (4.16)$$

While in the lattice models the Lipatov asymptotics only represent the logarithmic contributions of the zeroth and first orders, here we get 'extra' logarithms: formally we have $K = 0, 1, \ldots, \infty$ in Eqn (4.16), whereas $K \leqslant N$ in Eqn (4.11). This is the result of the rapid decrease of $A_N^K$ with increasing $K$, and the limited accuracy ($\sim 1/N$) of the principal asymptotics. Because of this, the result (4.16) can be only trusted when $K$ is not large.

The calculation of $A_N^K$ with $K \sim N$ is based on the renormalizability of the theory. Let us explain the basic idea. Let $F$ be a certain observable quantity. When we calculate it formally using the perturbation theory, it is a function of the bare charge $g_0$ and the cutoff parameter $\Lambda$ that has to be introduced for eliminating the divergences. The renormalizability of the theory means that the renormalized charge $g$ can be defined in such a way that the variable $F$ as a function of $g$ does not diverge, and at $\Lambda \to \infty$ tends to a finite limit,

$$F(g_0, \Lambda) = F_R(g). \qquad (4.17)$$

The quantities that are not directly observable (like the Green's function) may be renormalizable in a more complicated fashion; as a matter of fact, for all technical purposes we may confine ourselves to the category of quantities that are renormalizable in the multiplicative way,

$$F(g_0, \Lambda; p_i, \ldots) = Z(g_0, \Lambda) F_R(g; p_i, \ldots). \qquad (4.18)$$

In other words, from the quantity $F$ (which depends on the momenta $p_i$ and other variables) we separate out the diverging $Z$ factor. Since $F_R$ does not depend on $\Lambda$, we obtain

$$\frac{\mathrm{d} F_R}{\mathrm{d} \ln \Lambda} = 0. \qquad (4.19)$$

Substituting here $F_R$ from Eqn (4.18) and expressing the total derivative in terms of partial derivatives, we get the Callan–Symanzik equation

$$\left[ \frac{\partial}{\partial \ln \Lambda} + W(g_0) \frac{\partial}{\partial g_0} + V(g_0) \right] F(g_0, \Lambda; p_i, \ldots) = 0. \qquad (4.20)$$

Defined formally, the functions $W$ and $V$ depend on $\Lambda$. The actual absence of such a dependence may be proved taking advantage of the fact that the $Z$ factors for different $F$ are not independent, and can be expressed one via another [46]. Therefore, equation (4.20) contains much more than Eqn (4.19), reflecting the deep functional relations caused by the renormalizability: it is equivalent to the assumption of the existence of an exact renormalization group in Wilson's method [9, 10].

In the case under consideration multiplicative renormalizability is exhibited by the quantity $\varkappa^2 + \Sigma(0, \varkappa) - \Sigma(0, 0)$ [38]; it differs from the series in Eqn (4.11) only by a trivial term. Its substitution into Eqn (4.20) yields a set of equations in $A_N^K$, which allow the coefficients with $K > 0$ to be found given $A_N^0$. Hence, it becomes clear how powerful the property of renormalizability is: it dramatically reduces the uncertainty in the selection of coefficients $A_N^K$ in expansion (4.11). The coefficients $A_N^0$ are well enough represented by Lipatov's asymptotics [see Eqn (4.16)] and may serve as the boundary condition for this set of equations. This allows all $A_N^K$ with $N \gg 1$ to be analysed.

There is a certain 'gap' in the information regarding the coefficients $A_N^0$: for small $N$ they are calculated by the perturbation theory, for $N \gg 1$ by the Lipatov method, but for $N \sim 1$ they are, strictly speaking, not known, although it is possible to compute them by interpolation to an accuracy of a few percent (see examples in Refs [58, 62]). Accordingly, in the $(N, K)$ plane we have a region of nonuniversality $I$ (Fig. 7), in which the behavior of the coefficients $A_N^K$ depends considerably on the actual values of $A_N^0$ with $N \sim 1$. The regions above and below region $I$ are governed, respectively, by the trivial coefficient $A_0^0 = 1$ and the Lipatov asymptotics for $A_N^0$. The thick dashed line in Fig. 7 indicates the saddle-point values $K$ for $N = $ const under the assumption that $|g_0| \ll 1$, $g_0 \ln \Lambda/\varkappa \sim 1$; as $N$ decreases, the saddle disappears and the dominating role goes to the parquet coefficients $A_N^N$, lying on the principal diagonal. An important contribution to the sum (4.11) comes from regions $II$ and $III$ adjacent to the dashed saddle line: region $II$ gives the 'quasi-parquet' contribution, which is determined by the coefficients $A_N^{N-K}$ with $K \sim 1$ [see Eqn (4.13)] and only differs from the parquet contribution in that $g_0$ is replaced by the modified value $g_1$; region $III$ gives the nonperturbative contribution discussed in detail above. The result is similar to Eqn (4.14), the only difference being that the dependence of the nonperturbative term $i\Gamma_0(\varkappa)$ on $\varkappa$ is strong.

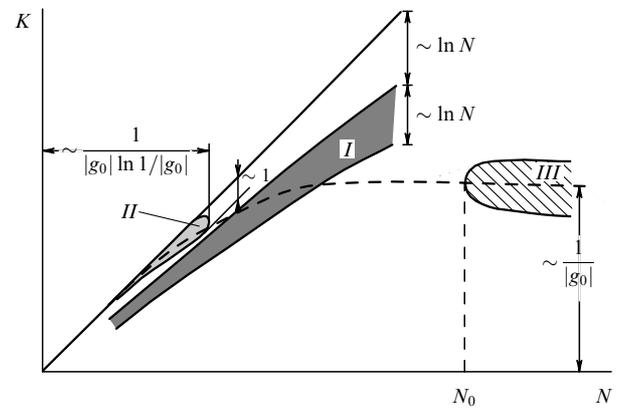

**Figure 7.** Regions in the $(N, K)$ plane important in the analysis of sum (4.11): $I$, region of nonuniversality, which requires knowing the coefficients $A_N^0$ with $N \sim 1$; the thick dashed line shows the saddle-point values of $K$ at $N = $ const; at small $N$, the parquet coefficients belonging to the main diagonal dominate; $II$, region giving the quasi-parquet contribution to sum (4.11); $III$, region giving the nonperturbative contribution into the sum (4.11) ($N_0$ may be selected as large as desired).

The spurious pole in Eqn (3.3) is avoided in such a way that $g \lesssim 1/\ln \ln(1/g_0)$, and the maximum value of the effective interaction is logarithmically small. This smallness is essential for obtaining closed results. The size of region $II$ in Fig. 7 is indicated for the smallest value of the pole denominator in Eqn (3.3); if we hypothetically further reduce this minimum value, region $II$ broadens and overlaps the nonuniversality region $I$. A quantitative description of the latter requires knowing the coefficients of expansion of the Gell-Mann–Low function at $N \sim 1$. This is exactly the information that is necessary for a complete reconstruction of the Gell-Mann–Low function, and is the focus of the entire scope of the tight-binding problems.

Let us demonstrate that the exponent $\alpha$ in Eqn (2.13) can be found from the condition of renormalizability of the plateau contribution. The contribution of the plateau to the damping $\Gamma$, which depends on $\Lambda$ and the bare values $\varkappa_0$ and $g_0$, becomes a function of only $\varkappa$ and $g$ upon transition to the renormalized values. From considerations of dimensionality we have $\Gamma = \varkappa^2 f(g)$, where $f(g)$ is mainly determined by the exponential $\exp(-1/ag)$ to ensure agreement with the result of the method of optimal fluctuation at $E \to -\infty$, when $g \approx g_0$. With due account for Eqn (3.3), we obtain

$$\Gamma \sim \varkappa^2 \exp\left(-\frac{1}{ag_0} - \frac{W_2}{a}\ln\frac{\Lambda}{\varkappa}\right)$$
$$\sim \Lambda^2 \left(\frac{\Lambda^2}{\varkappa^2}\right)^{-W_2/2a-1} \exp\left(-\frac{1}{ag_0}\right), \quad (4.21)$$

which, given that $J \sim \Lambda^2$, $\varkappa^2 = |E|$, $a = -3/8\pi^2$, reproduces the second term in Eqn (2.14)†; this value of $a$ is obtained by the method described in Ref. [20] using the known instanton solution for $d = 4$ [58].

### 4.4 Transition to the $(4-\epsilon)$-dimensional theory
For $d = 4 - \epsilon$, the expansion for the self-energy is similar to Eqn (4.11) with the replacement

$$\ln\left(\frac{\Lambda}{\varkappa}\right) \longrightarrow \frac{(\Lambda/\varkappa)^\epsilon - 1}{\epsilon}. \quad (4.22)$$

The coefficients $A_N^K$ become functions of $\epsilon$ and their calculation is entirely similar to the four-dimensional case, i.e., is based on the renormalizability of the theory with Lipatov's asymptotics as boundary conditions. A consistent presentation of the structure of the theory is given in Section 6.

The typical behavior of the density of states $v(E)$ for the four types of models under consideration is shown in Fig. 8.

## 5. Higher order perturbation theory

In this section we are going to defend a thesis which, in our opinion, has not quite been assimilated by the scientific community: *in the higher orders it is possible to calculate everything.*

### 5.1 Statistical method
For standard models of the $\varphi^4$ type, the perturbation-theory expansion diverges factorially, which is related to the factorially large number of diagrams of the same order. The large number of diagrams allows hope for the success of statistical methods, since the law of large numbers at work should lead to simple limiting distributions characterized by a small number of parameters. Such a program was realized in Ref. [36] for the diagram series for self-energy.

Given that the diagram of the $(N+1)$st order can be obtained from the diagram of the $N$th order by 'hitching' the additional impurity stipple, the linkage between the diagrams of the $(N+1)$st and $N$th order may be expressed in the form of the recurrence relation

$$\begin{vmatrix} \Sigma'_{N+1} \\ \Sigma''_{N+1} \end{vmatrix} = \begin{vmatrix} A_N & B_N \\ C_N & D_N \end{vmatrix} \begin{vmatrix} \Sigma'_N \\ \Sigma''_N \end{vmatrix}, \quad (5.1)$$

where $\Sigma'_N$ and $\Sigma''_N$ are the real and imaginary parts of some diagram of the $N$th order. Evaluating the coefficients $A_N$, $B_N$, $C_N$, and $D_N$ by order of magnitude, we pick out the dimensional parameters. Then the hitched stipples are divided into 'classes', so that each class is characterized by the values of $A_N$, $B_N$, $C_N$, and $D_N$ falling into a certain small interval. The diagrams obtained from the first diagram in Fig. 1b by repeated hitching of stipples of one and the same class constitute, as $N \to \infty$, an infinitesimal fraction of the total number of diagrams. In a typical diagram of $N$th order the class of stipple at each step is selected at random. This allows us to go over to a statistical description and treat $A_N$, $B_N$, $C_N$, and $D_N$ as random quantities.

† Results of the type of Eqn (2.14) involve the renormalized energy $E$ [43].

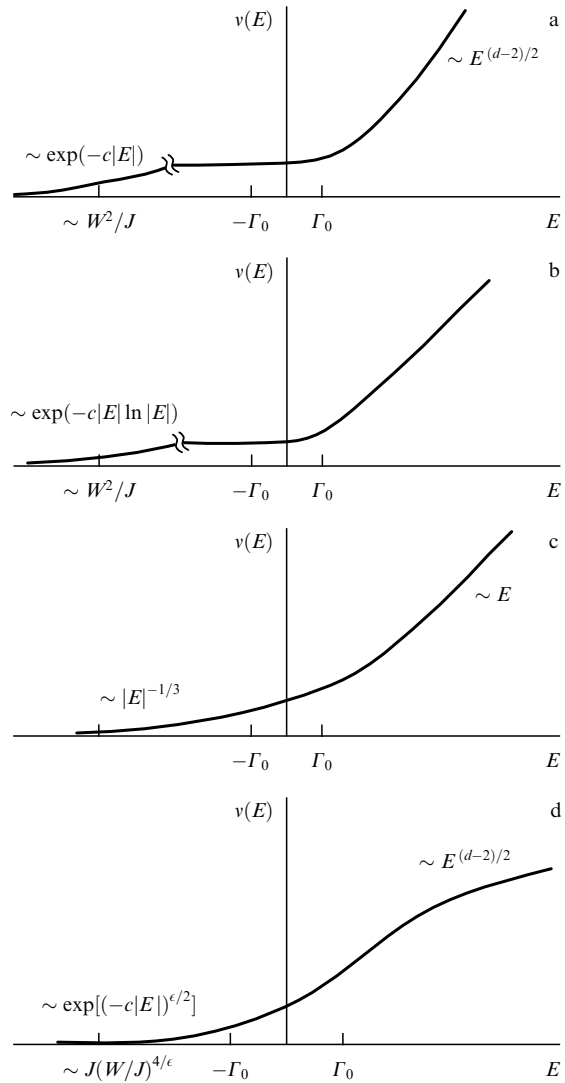

**Figure 8.** Density of states $v(E)$ for four types of models: (a) $d > 4$; (b) $d = 4$, lattice models; (c) $d = 4$, renormalizable models; (d) $d = 4 - \epsilon$. The scale $\Gamma_0$ is exponentially small as a function of the parameter $1/g_0$ (a, b, c), or $1/\epsilon$ (d).

Analyzing equation (5.1) with random parameters, one may show that the contribution from an individual diagram of $N$th order is a highly fluctuative quantity, but the total contribution of the $N$th order is self-averaging. As a result, the common term of the series for large $N$ is found, and the functional form of the fluctuation tail of the density of states is reconstructed, which earlier could only have been obtained by the instanton method. Thus, the statistical method of Ref. [36] demonstrates that it is possible in principle to reproduce the instanton results by the diagram technique, which has been questioned even in serious studies.

### 5.2 Lipatov method
The canonical procedure for calculating the remote terms in the perturbation expansion is the Lipatov method [58] as described above (Section 4.1). This method allows the remote coefficients of expansion in the coupling constant $g$ to be calculated for practically any functional integral. While the exact calculation of a functional integral is, as a rule, not possible, its remote coefficients can be calculated in the saddle-point approximation, which can always be implemented.

For the problem in question we are interested in the integrals

$$Z_M(\alpha_1, x_1, \ldots, \alpha_M, x_M)$$
$$= \int D\varphi \, \varphi_{\alpha_1}(x_1) \ldots \varphi_{\alpha_M}(x_M) \exp(-H\{g, \varphi\}), \quad (5.2)$$

where

$$H\{g, \varphi\} = \int d^d x \left\{ \frac{1}{2} \sum_\alpha [\nabla \varphi_\alpha(x)]^2 + \frac{1}{2} \varkappa^2 \sum_\alpha \varphi_\alpha(x)^2 + \frac{1}{4} g \left[ \sum_\alpha \varphi_\alpha(x)^2 \right]^2 \right\}. \quad (5.3)$$

Following the general scheme of Lipatov's method (Section 4.1) and replacing $F(g)$ in Eqn (4.5) with $Z_M$, we get

$$[Z_M]_{N-1} = \int \frac{dg}{2\pi i} \int D\varphi \, \varphi_{\alpha_1}(x_1) \ldots \varphi_{\alpha_M}(x_M)$$
$$\times \exp(-H\{g, \varphi\} - N \ln g). \quad (5.4)$$

The saddle point is defined by the condition

$$\begin{pmatrix} \dfrac{\partial}{\partial g} \\ \dfrac{\delta}{\delta \varphi} \end{pmatrix} (-N \ln g - H\{g, \varphi\}) = 0. \quad (5.5)$$

Seeking the classical solution (instanton) in the form $\varphi_\alpha^c(x) = \varphi_c(x) u_\alpha$, where $u_\alpha$ is a component of the unit vector **u**, from Eqn (5.5) we get

$$N g_c^{-1} = -\frac{1}{4} \int d^d x \, \varphi_c(x)^4, \quad (5.6)$$

$$-\Delta \varphi_c(x) + \varkappa^2 \varphi_c(x) + g_c \varphi_c(x)^3 = 0, \quad (5.7)$$

whence it follows that $g_c < 0$, and $\varphi_c(x)$ is sought among the class of functions decreasing at infinity to ensure convergence of the integral in Eqn (5.6).

Considering small deviations from the saddle point

$$g = g_c + \delta g, \quad \varphi_\alpha(x) = \varphi_c(x) u_\alpha + \delta \varphi_\alpha(x), \quad (5.8)$$

it is easy to prove the validity of estimates

$$g_c \sim N^{-1}, \quad \delta g \sim N^{-3/2}, \quad \varphi_c \sim N^{1/2}, \quad \delta \varphi \sim 1, \quad (5.9)$$

which confirm that the saddle-point method works for large $N$.

### 5.3 Algebra of factorial series

In general, knowledge of the coefficients in the expansions of functional integrals is by itself not sufficient, since it is usually necessary to perform certain manipulations with these integrals: find the $M$-point Green's function defined as the ratio of two integrals,

$$G_M(\alpha_1, x_1, \ldots, \alpha_M, x_M) = Z_0^{-1} Z_M(\alpha_1, x_1, \ldots, \alpha_M, x_M), \quad (5.10)$$

go over to the self-energy or the vertex part, calculate the free energy as the logarithm of the statistical integral, etc. If, however, we are only interested in the remote coefficients of the expansion, then a straightforward algebra is available for the factorial series that allows the latter to be manipulated as easily as finite expressions.

Let

$$S_A = A_0 + A_1 g + \ldots + A_N g^N + \ldots,$$
$$S_B = B_0 + B_1 g + \ldots + B_N g^N + \ldots \quad (5.11)$$

be two factorial series, so that $A_N, B_N \sim N!$. Straightforward multiplication of the series yields

$$S_A S_B = A_0 B_0 + \ldots + g^N (A_0 B_N + A_1 B_{N-1}$$
$$+ A_2 B_{N-2} + \ldots + A_N B_0) + \ldots \quad (5.12)$$

Since the series are factorial, we have $B_{N-1} \sim B_N/N$, $B_{N-2} \sim B_N/N^2 \ldots$, and, similarly, $A_{N-1} \sim A_N/N$, $A_{N-2} \sim A_N/N^2 \ldots$ This allows us to retain only the first and last terms in parentheses in Eqn (5.12),

$$S_A S_B = A_0 B_0 + \ldots + g^N (A_0 B_N + A_N B_0) + \ldots, \quad (5.13)$$

if the coefficients $A_N$ and $B_N$ have the same rate of growth. If the coefficients of one of the series, $S_A$ for example, grow faster (owing to the slower corrections to the principal factorial dependence), then only the second term in parentheses is retained in Eqn (5.13), and the product $S_A S_B$ can be written as $B_0 S_A$, if we only take into account the first term and the remote terms in the expansion.

Using Eqn (5.13) for the $n$-fold multiplication of the series $S_A$ by itself, we get the formula for raising $S_A$ to a natural power:

$$(S_A)^n = A_0^n + \ldots + n A_0^{n-1} A_N g^N + \ldots \quad (5.14)$$

Making use of this formula, it is easy to get the result for any regular function $f(x)$

$$f(S_A) = f(A_0) + \ldots + g^N f'(A_0) A_N + \ldots \quad (5.15)$$

and extend the result (5.14) to negative and fractional powers.

### 5.4 Problems to be solved

In principle, the idea of Lipatov's method is simple enough, since the expansion in the neighborhood of the saddle point gives rise to a Gaussian integral which can always be calculated:

$$\int dt_1 \int dt_2 \ldots \int dt_N \exp\left(-\sum_{i,j=1}^N A_{i,j} t_i t_j\right) = \frac{(2\pi)^{N/2}}{(\det A)^{1/2}}. \quad (5.16)$$

Practical calculations, however, are rather cumbersome because they run into a number of technical problems. Without going into the details, which are discussed at length in Refs [37–39], we shall discuss the main problems which have to be solved.

**5.4.1 Existence of instantons.** The question of the existence of instantons is nontrivial [44]: the solutions of Eqn (5.7) exist at $d < 4$ for $\varkappa^2 > 0$, but not at $d > 4$. At $d = 4$, the equation admits solutions for $\varkappa = 0$, but not for $\varkappa$ other than zero. The physical meaning of this peculiar situation becomes clear from the estimates made by the method of optimal fluctuation† (Section 2). With $d > 4$ and $d = 4$, $E < 0$, the minimum of the function $S(E, R)$ in Eqn (2.8) is achieved at the boundary of the domain of definition, and its derivative with respect to $R$ does not go to zero; accordingly, the variation of $H\{g, \varphi\}$ with respect to $\varphi$ which leads to Eqn (5.7) is nonzero. This problem is resolved in different ways for $d > 4$ and $d = 4$.

At $d > 4$, the theory is not renormalizable, and the continuum model is logically not closed: it must be extended to small distances. This can be accomplished by passing to the lattice model [then $-\Delta = \hat{p}^2$ in Eqn (5.7) is replaced by $\epsilon(\hat{p})$], or by introducing the finite correlation radius of the random potential; then the term $\varphi^4$ in Eqn (5.3) becomes

$$\frac{1}{4} g \sum_\alpha \int d^d x \int d^d y \, \varphi_\alpha(x)^2 B(x-y) \varphi_\alpha(y)^2 \,. \quad (5.17)$$

After such modifications, instantons arise and are localized on the relevant minimum length scale. The same treatment should be applied to the nonrenormalizable class of models at $d = 4$.

In the case of four-dimensional renormalizable models, the problem can be solved without such modifications. The idea consists in minimizing $H\{g, \varphi\}$ with a fixed instanton radius $R$ and calculating the integral with respect to $R$, which is essentially non-Gaussian. The same procedure is used for $d = 4 - \epsilon$; in this case the instanton formally exists, but the direction corresponding to the change in the instanton radius is 'soft', and the saddle-point approximation is not sufficient‡.

Let us define the 'center' $x_0$ of the function $\boldsymbol{\varphi}(x)$ and its 'radius' $R$ by the conditions§

$$\int d^d x \, |\boldsymbol{\varphi}(x)|^4 (x-x_0)_\mu = 0, \quad \mu = 1, 2, \ldots, d, \quad (5.18)$$

$$\int d^d x \, |\boldsymbol{\varphi}(x)|^4 \ln \frac{(x-x_0)^2}{R^2} = 0 \quad (5.19)$$

and include the expansion of unity into the integrand in Eqn (5.4)

$$1 = \int d^d x \, |\boldsymbol{\varphi}(x)|^4 \int_0^\infty d\ln R^2 \delta\left[-\int d^d x \, |\boldsymbol{\varphi}(x)|^4 \ln\left(\frac{x-x_0}{R}\right)^2\right]. \quad (5.20)$$

Now we change the order of integration with respect to $R$ and $\varphi$; then the $\delta$ function in Eqn (5.20) limits integration with respect to $\varphi$ by condition (5.19) (assumption of constant radius $R$), and integration with respect to $R$ is explicit.

---

† The method of optimal fluctuation corresponds to the replica limit $n = 0$; with $n \neq 0$, similar results can be obtained by characterizing the instanton with two parameters (amplitude and radius) and carrying out a variation evaluation of action.

‡ More precisely, the saddle-point approximation is applicable when $N\epsilon \gg 1$, whereas the calculations are carried out under a weaker assumption $N \gg 1$.

§ Obviously, the result should not depend on the way in which $x_0$ and $R$ are defined. One may prove that if $|\boldsymbol{\varphi}(x)|^4$ is replaced by $|\boldsymbol{\varphi}(x)|^\alpha$ in Eqns (5.18) and (5.19), the result does not contain the parameter $\alpha$; we have not analyzed this issue in further details. Similar remarks apply to Eqn (5.30) below.

If expansion (5.8) is carried out in the neighborhood of an arbitrary function $\varphi_c(x)$, then the exponential (5.4) exhibits terms linear in $\delta\varphi$. They only vanish for the 'true' instanton satisfying Eqn (5.7). In the case we are interested in, however, there are no 'true' instantons because Eqn (5.7) is insolvable. Let us select the function $\varphi_c(x)$ from the condition of minimum $H\{g_c, \varphi\}$ under the additional assumption (5.19) which fixes the instanton radius. After the change of variables this brings us to the equation

$$-\Delta \varphi_c(x) + \varkappa_R^2 \varphi_c(x) + g_c \varphi_c(x)^3 + \mu \varphi_c(x)^3 \ln x^2 = 0, \quad (5.21)$$

where $\varkappa_R = \varkappa R$, and $\mu$ is the Lagrange multiplier [57]. Such a selection for the instanton has a 'miraculous' property: the argument of the $\delta$ function (5.20) features a combination which exactly coincides with the terms linear in $\delta\varphi$ in the exponential (5.4), so that these terms are totally eliminated.

**5.4.2. Zero and soft modes.** If matrix $A$ in Eqn (5.16) has a zero eigenvalue $\lambda_0$, integral (5.16) diverges. Such a divergence is actually fictitious: this is easily demonstrated by taking a finite eigenvalue $\lambda_0$ and reducing it to zero. Such a passage to the limit reveals a 'soft' direction, along which the fluctuations grow indefinitely. The divergence of integral (5.16) only implies that the Gaussian approximation does not work, and exact integration in the 'soft' direction is required.

As a matter of fact, the zero modes arise in all situations of interest, and their occurrence can be anticipated. Splitting $\delta\varphi_\alpha(x)$ in Eqn (5.8) into the longitudinal and transverse components,

$$\delta\varphi_\alpha(x) = \delta\varphi_L(x) u_\alpha + \delta\varphi_\alpha^T(x), \quad \boldsymbol{\delta\varphi}^T \perp \mathbf{u}, \quad (5.22)$$

it is easy to see that exponential (5.4) contains the expression

$$N\left(\frac{\delta g}{g_c}\right)^2 - 2\delta g \int d^d x \, \varphi_c(x)^3 \delta\varphi_L(x)$$

$$- \int d^d x \, \delta\varphi_L(x) \hat{M}_L \delta\varphi_L(x) - \sum_\alpha \int d^d x \, \delta\varphi_\alpha^T(x) \hat{M}_T \delta\varphi_\alpha^T(x), \quad (5.23)$$

where we have introduced the operators $\hat{M}_L$ and $\hat{M}_T$ that play an important role in the instanton calculations [20],

$$\hat{M}_L = -\Delta + \varkappa^2 - 3\phi_c(x)^2, \quad \hat{M}_T = -\Delta + \varkappa^2 - \phi_c(x)^2, \quad (5.24)$$

and changed variables

$$\varphi_c(x) = (-g_c)^{-1/2} \phi_c(x), \quad (5.25)$$

to eliminate $g_c$ from Eqn (5.7).

In the ordinary situation, when 'true' instantons satisfying Eqn (5.7) exist, one may demonstrate the existence of the zero translation and rotation modes [20]. Indeed, if $\varphi_c(x)$ is a solution of Eqn (5.7), then $\varphi_c(x - x_0)$ is also a solution; passing to the limit of small $x_0$, we find that operator $\hat{M}_L$ has $d$ zero modes

$$e_\mu^L(x) = \mathrm{const} \times \frac{\partial \phi_c(x)}{\partial x_\mu}, \quad \mu = 1, 2, \ldots, d. \quad (5.26)$$

Similarly, if $\varphi_\alpha^c(x) = \varphi_c(x) u_\alpha$ is an instanton, then $\varphi_c(x) u'_\alpha$ is also an instanton. Because of this, operator $\hat{M}_T$ has a zero mode

$$e_0^T(x) = \text{const} \times \phi_c(x), \qquad (5.27)$$

which immediately follows from Eqn (5.7).

In the 'massless' four-dimensional theory ($d = 4$, $\varkappa = 0$), equation (5.7) after replacement (5.25) admits a solution (see Ref. [58])

$$\phi_c(x) = \frac{2\sqrt{2}\, R}{x^2 + R^2}, \qquad (5.28)$$

where the parameter $R$ (the instanton radius) is arbitrary in accordance with the plateau in the method of optimal fluctuation (Fig. 4b); the uncertainty of parameter $R$ results in that $\hat{M}_L$ displays yet another zero mode, the dilatation mode [58, 61]

$$e_0^L(x) = \text{const} \times \frac{\partial \phi_c(x)}{\partial R}. \qquad (5.29)$$

In fact, it is this mode that leads to the insolvability of the four-dimensional equation (5.7) for finite $\varkappa$: the inclusion of $\varkappa^2$ in the perturbation theory leads to the equation $\hat{M}_L \delta\varphi = -\varkappa^2 \varphi_c(x)$, whose right-hand side is not orthogonal to $e_0^L(x)$.

In the 'massive' four-dimensional theory ($d = 4$, $\varkappa^2 > 0$), and at $d = 4 - \epsilon$, when the instanton is defined by Eqn (5.21), all the above modes cease to be zero modes but are still soft, since the Lagrange multiplier $\mu$ contains a small parameter [38, 39]. Treatment of the translation and rotation modes is performed in a way completely similar to the integration with respect to the instanton radius (which corresponds to the dilatation mode). In addition to the 'center' and 'radius' of the function $\boldsymbol{\varphi}(x)$, we define its 'orientation' $\mathbf{v}$ as the unit vector fixed by the condition

$$\mathbf{v}\{\varphi\} \| \int d^d x \, |\boldsymbol{\varphi}(x)|^3 \boldsymbol{\varphi}(x). \qquad (5.30)$$

Along with Eqn (5.20), we introduce two expansions of unity into the integrand in Eqn (5.4)

$$1 = \left( \int d^d x \, |\boldsymbol{\varphi}(x)|^4 \right)^d \int d^d x_0 \prod_{\mu=1}^{d} \delta\left( -\int d^d x \, |\boldsymbol{\varphi}(x)|^4 (x - x_0)_\mu \right), \qquad (5.31)$$

$$1 = \int d^n u \, \delta(\mathbf{u} - \mathbf{v}\{\varphi\}), \qquad (5.32)$$

and carry out integration with respect to $x_0$ and $\mathbf{u}$ from integral with respect to $\varphi$. Then the functional integration is performed with a fixed center and orientation of the instanton, and can be carried out in the saddle-point approximation.

**5.4.3 Practical calculation of determinants.** After the integration with respect to $\varphi$, Eqn (5.4) will contain the determinants of operators $\hat{M}_L$ and $\hat{M}_T$ with the contributions of zero modes eliminated; the spectrum of these operators contains a continuous component, and the calculation of determinants requires carrying out quantization in a large but finite volume with subsequent passage to the thermodynamic limit. However, it is not possible, as a rule, to calculate the eigenvalues of $\hat{M}_L$ and $\hat{M}_T$ analytically, and the numerical techniques do not work well in such a situation. A technique for dealing with this difficulty was proposed by Brezin and Parisi [62]. Introducing the notation

$$D(z) = \det \hat{R}(z), \qquad \hat{R}(z) = 1 - \frac{3z\phi_c(x)^2}{\hat{p}^2 + \varkappa^2} \qquad (5.33)$$

and taking advantage of the fact that the determinant of a product is equal to the product of the determinants, we get

$$\det \hat{M}_L = D(1) \det\{\hat{p}^2 + \varkappa^2\},$$

$$\det \hat{M}_T = D\left(\frac{1}{3}\right) \det\{\hat{p}^2 + \varkappa^2\}. \qquad (5.34)$$

The determinant of the operator $\hat{p}^2 + \varkappa^2$ is defined by the functional integral for the problem without interaction, and its calculation is straightforward. The aim of these manipulations is to use the fact that the spectrum of the operator $\hat{R}(z)$ is purely discrete: its lowest eigenvalues can be found numerically, and the higher ones are approximated by simple asymptotics [see (5.39)]. It is easy to prove that

$$D(z) = \prod_s \left(1 - \frac{z}{\mu_s}\right), \qquad (5.35)$$

where $\mu_s$ are the eigenvalues of the equation

$$[\hat{p}^2 + \varkappa^2 - 3\mu_s \phi_c(x)^2] e_s(x) = 0. \qquad (5.36)$$

By virtue of Eqn (5.35), $\det \hat{M}_L$ and $\det \hat{M}_T$ are expressed in terms of one sequence $\mu_s$.

In order to eliminate the contribution of the zero (or soft) modes from the determinant, one must know the law according to which the relevant eigenvalues $\lambda_s(z)$ of the operator $-\Delta + \varkappa^2 - 3z\phi_c(x)^2$ tend to zero when $z \to 1$ or $z \to 1/3$. The perturbation theory yields the following results corresponding to the translation, rotation and (at $d = 4$) dilation modes:

$$\lambda_\mu(z) = I_6 \left[ 4 \int d^d x \left( \frac{\partial \phi_c(x)}{\partial x_\mu} \right)^2 \right]^{-1} (z - 1), \quad z \to 1,$$

$$\lambda_0(z) = \frac{I_4}{I_2} (1 - 3z), \quad z \to \frac{1}{3}, \qquad (5.37)$$

$$\lambda_0(z) = 3J \left[ \int d^4 x \left( \frac{\partial \phi_c(x)}{\partial R} \right)^2 \right]^{-1} (z - 1), \quad z \to 1,$$

where we use the notation for the integrals of the instanton solution

$$I_p = \int d^d x \, \phi_c(x)^p, \qquad J = \int d^4 x \, \phi_c(x)^2 \left( \frac{\partial \phi_c(x)}{\partial R} \right)^2 \bigg|_{R=1}. \qquad (5.38)$$

**5.4.4 Divergences of determinants.** The method of Brezin and Parisi exposes divergences in the determinants, the elimination of which reveals a linkage with the general problems of renormalizability. Let us find the asymptotics for $\mu_s$ with large $s$, counting the number of electrons whose energy is less than $-\varkappa^2$ in the quasi-classical (with large $\mu$) potential $-3\mu\phi_c(x)^2$, which can be accomplished using the Thomas–

Fermi method [62]. The result is

$$\mu_s \sim s^v, \tag{5.39}$$

where $v = (2d-4)/d$ in the lattice models for $d \geq 4$ [37] and $v = 2/d$ in the continuum models for $d \leq 4$ [39]. Representing Eqn (5.35) in the form

$$D(z) = \exp\left(-z\sum_s \mu_s^{-1} - \frac{1}{2}z^2\sum_s \mu_s^{-2} - \frac{1}{3}z^3\sum_s \mu_s^{-3} - \ldots\right), \tag{5.40}$$

we see that with $d > 4$ the sequence $\mu_s$ grows faster than $s^{1+\delta}$, and there are no divergences in $D(z)$. With $d < 4$, and with $d = 4$ in the lattice models, the first sum in Eqn (5.40) diverges. In the continuum model with $d = 4$, the first two sums diverge. These divergences can be set apart explicitly with the aid of the sum rules obtained by calculating $\ln D(z)$ for small $z$ by perturbation theory based on the definition (5.33) [38, 39, 62]:

$$\sum_s \frac{1}{\mu_s} = 3I_2 \int_0^\Lambda \frac{d^d p}{(2\pi)^d} \frac{1}{p^2 + \varkappa^2},$$

$$\sum_s \frac{1}{\mu_s^2} = 9 \int_0^\Lambda \frac{d^d p}{(2\pi)^d} \int_0^\Lambda \frac{d^d q}{(2\pi)^d} \frac{\langle \phi_c^2 \rangle_q \langle \phi_c^2 \rangle_{-q}}{(p^2 + \varkappa^2)[(p+q)^2 + \varkappa^2]}, \tag{5.41}$$

where $\langle f \rangle_q$ is the Fourier component of the function $f(x)$. For $d = 4 - \epsilon$, the second sum becomes

$$\sum_s \frac{1}{\mu_s^2} \approx 9K_d I_4 \frac{1 - \Lambda^{-\epsilon}}{\epsilon} + 12\left(\frac{1}{3} + C - \ln 2\right). \tag{5.42}$$

The divergence of the first sum is eliminated as a result of the renormalization of the quantity $\varkappa$. If its bare value is $\varkappa_0$, then the initial Hamiltonian can be represented as the sum of Hamiltonian (5.3) with renormalized $\varkappa$ and the counter-term

$$\Delta H\{\varphi\} = \frac{1}{2}\sum_x (\varkappa_0^2 - \varkappa^2)\sum_\alpha \varphi_\alpha(x)^2, \tag{5.43}$$

where the renormalization of $\varkappa$ is sought in the form of the diagram expansion

$$\varkappa_0^2 - \varkappa^2 = C_1 g + C_2 g^2 + \ldots \tag{5.44}$$

Constructing the instanton based on Hamiltonian (5.3) and estimates (5.9) for the values at the saddle point and the rms fluctuations in its neighborhood, we find that for the calculations up to zero-order accuracy with respect to $N$, one must include in Eqn (5.44) only the first term of the expansion, and substitute into Eqn (5.43) the values of $g$ and $\varphi_\alpha(x)$ at the saddle point. Calculating $\varkappa_0^2 - \varkappa^2$ in the one-loop approximation [which defines $C_1$ in Eqn (5.44)], we find that the counterterm (5.43) gives rise to the extra factor

$$\exp\left(\frac{1}{2}g_c(n+2)\int \frac{d^d p}{(2\pi)^d}\frac{1}{p^2 + \varkappa^2}I_2\right), \tag{5.45}$$

which exactly cancels the diverging part of the determinants. The divergence of the second sum in Eqn (5.41) at $d = 4$ can be eliminated by renormalizing the charge†; we, however, are interested in an expansion of the type of Eqn (4.2), which contains the bare charge $g_0$ and the logarithmic divergences in explicit form. The final result is expressed in terms of the renormalized determinants, which are defined as

$$D_R(z) = \prod_s \left(1 - \frac{z}{\mu_s}\right)\exp\left(\frac{z}{\mu_s}\right) \tag{5.46}$$

or

$$D_R(z) = \prod_s \left(1 - \frac{z}{\mu_s}\right)\exp\left(\frac{z}{\mu_s} + \frac{z^2}{2\mu_s^2}\right). \tag{5.47}$$

The last expression is used in the continuum model at $d = 4$; it is also convenient for expressing the results for $d = 4 - \epsilon$. The products (5.46) or (5.47) at $z = 1$ and $z = 1/3$ with eliminated zero multipliers corresponding to the contributions from zero modes are denoted by $\bar{D}_R(1)$ and $\bar{D}_R(1/3)$.

### 5.5 Summary of instanton results

The Lipatov method yields results for the coefficients of expansion of arbitrary $M$-point Green's functions [37–39, 61, 62] with an arbitrary number $n$ of field components $\varphi$, from which one can pass to the self-energy and vertex parts, and then to the quantities of direct physical interest: the Gell-Mann–Low function [58], other scaling functions [61, 62], the density of states in the fluctuation tail [37–39, 43], etc.

We are interested in the self-energy of a disordered system ($n = 0$). In the lattice models with $d \geq 4$, we have [37]

$$[\Sigma(p,\varkappa)]_N = \frac{\langle \phi_c^3 \rangle_p \langle \phi_c^3 \rangle_{-p}}{(2\pi)^{1/2}}\left(\frac{\bar{D}_R(1/3)}{|\bar{D}_R(1)|}\right)^{1/2}$$

$$\times \frac{2}{I_4(\varkappa^2)}\Gamma\left(N + \frac{1}{2}\right)\left(-\frac{4}{I_4(\varkappa^2)}\right)^N, \tag{5.48}$$

where

$$I_4(\varkappa^2) = I_4(0) + 2I_2(0)\varkappa^2, \quad d > 4, \tag{5.49}$$

$$I_4(\varkappa^2) = I_4(0) + 2I_3(0)^2 K_4 \varkappa^2 \ln\frac{\Lambda}{\varkappa} + O(\varkappa^2), \quad d = 4. \tag{5.50}$$

Here the lattice instanton is defined by Eqn (5.7) with $-\Delta = \hat{p}^2 \to \epsilon(\hat{p})$, and the values of $I_p(\varkappa^2)$ are given by Eqn (5.38), where the integral with respect to $d^d x$ is replaced by the sum over the lattice sites. The translation modes in lattice models are gaplike, and there is no need to single them out from $D_R(1)$.

In the continuum model with $d = 4 - \epsilon$ ($0 \leq \epsilon \leq 1$), we have [38, 39]

$$[\Sigma(p,\varkappa)]_N = c_2\Gamma(N+b)a^N \int_0^\infty d\ln R^2 R^{-2}\langle\phi_c^3\rangle_{Rp}\langle\phi_c^3\rangle_{-Rp}$$

$$\times \exp\left[-Nf(\varkappa R) + N\epsilon \ln R + 2K_d I_4(\varkappa R)\frac{1 - (\Lambda R)^{-\epsilon}}{\epsilon}\right], \tag{5.51}$$

---

† This can be accomplished in a similar way using the most general form of counterterms (see Eqns (3.26)–(3.28) in Ref. [46]).

where

$$a = -3K_4, \quad b = \frac{d+2}{2}, \quad c_2 = c(3K_4)^{7/2},$$

$$I_4(x) = I_4 \exp[f(x)],$$

$$f(x) = -\frac{\epsilon}{2}(C + 2 + \ln \pi) - 3x^2\left(C + \frac{1}{2} + \ln \frac{x}{2}\right),$$

$$\langle \phi_c \rangle_p^3 = 8 \times 2^{1/2} \pi^2 p K_1(p), \quad (5.52)$$

$C$ is the Euler constant, $K_1(x)$ is the McDonald function, and the constant $c$ is defined as

$$c = \frac{2^{n-2} 3^{1/2}}{(2\pi)^{3+n/2}} \left(\frac{I_6}{4}\right)^2 J^{1/2} \exp\left[-\frac{3(n+4)}{4}\right.$$

$$\left. + \frac{n+8}{3}\left(-\ln 2 + C + \frac{1}{3}\right)\right]\left[-\bar{D}_R(1)\bar{D}_R^{n-1}\left(\frac{1}{3}\right)\right]^{-1/2}. \quad (5.53)$$

Here the integrals $I_p$ correspond to the four-dimensional massless theory and are defined from instanton (5.28) with $R = 1$:

$$I_3 = 4\sqrt{2}\,S_4, \quad I_4 = \frac{16}{3}S_4, \quad I_6 = \frac{64}{5}S_4,$$

$$J = \frac{16}{15}S_4; \quad S_4 = 2\pi^2. \quad (5.54)$$

In this case, the eigenvalues $\mu_s$ of Eqn (5.36) can be found exactly, and are equal to $s(s+1)/6$ with the degree of degeneracy $s(s+1)(2s+1)/6$ ($s = 1, 2, \ldots$). This allows the numerical values of the determinants $\bar{D}_R(1)$ and $\bar{D}_R(1/3)$ to be found:

$$-\bar{D}_R(1) \approx 578, \quad \bar{D}_R\left(\frac{1}{3}\right) \approx 0.872. \quad (5.55)$$

The result for $[\Sigma(p, \varkappa)]_N$ can also be found for $d < 4$ without assuming that $\epsilon$ is small. Then the dilation mode is considered on common grounds, and does not require singling out:

$$[\Sigma(p, \varkappa)]_N = \frac{1}{2(2\pi)^{(d+1)/2}}\left(\frac{I_6 - I_4}{I_4 d}\right)^{d/2}\left(\frac{4}{I_4}\right)^{(d+2)/2}$$

$$\times \left|\frac{\bar{D}_R(1/3)}{\bar{D}_R(1)}\right|^{1/2} \Gamma\left(N + \frac{d+1}{2}\right)$$

$$\times \left(-\frac{4}{I_4}\varkappa^{d-4}\right)^N \varkappa^2 \langle \phi_c^3 \rangle_{p/\varkappa} \langle \phi_c^3 \rangle_{-p/\varkappa}. \quad (5.56)$$

Here the instanton is defined by Eqn (5.7) with $\varkappa = 1$, and we have passed to the limit $\Lambda \to \infty$. For $d = 4 - \epsilon$, the applicability of the results (5.56) and (5.51) is essentially different: the saddle-point approximation was only used in the derivation of Eqn (5.51) in the integration with respect to the 'hard' variable, and the result is valid for $N \gg 1$, whereas Eqn (5.56) involved the use of the saddle-point approximation also for integrating with respect to the soft dilation mode, which is only justified if $N\epsilon \gg 1$. It is easy to prove [39] that in the latter case the results (5.56) and (5.51) coincide.

The traditional version of the instanton method yields the density of states of a disordered system only in the region of the fluctuation tail, when the density is determined by the sum of remote terms of the Green's function expansion,

$$v_{fl}(E) = \frac{1}{\pi} \operatorname{Im} \sum_{N=N_0}^{\infty} [G(x,x)]_N (g_0 - i0)^N, \quad N_0 \gg 1. \quad (5.57)$$

For $d < 4$, we obtain

$$v(E) = \frac{(4-d)2^{d-1}}{(2\pi)^{(d+1)/2}}\left(\frac{I_6 - I_4}{I_4 d}\right)^{d/2}\left|\frac{\bar{D}_R(1/3)}{\bar{D}_R(1)}\right|^{1/2}$$

$$\times |E|^{(d-2)/2}\left(\frac{I_4|E|^{(4-d)/2}}{2a_0^d W^2}\right)^{(d+1)/2} \exp\left(-\frac{I_4|E|^{(4-d)/2}}{2a_0^d W^2}\right),$$
$$(5.58)$$

which exhibits the same energy dependence as the result reported by Cardy [41]. Normalization with respect to the unperturbed density of states $v_0(E)$, and transition from the renormalized energy $E$ to the bare energy $E_B$ with biased zero (see Eqn (12) in Ref. [43]), yields the results of Brezin and Parisi [43]

$$\frac{v(E_B)}{v_0(-E_B)} = \left(\frac{I_6 - I_4}{3}\right)^{3/2}\left|I_4 \frac{\bar{D}_R(1/3)}{\bar{D}_R(1)}\right|^{1/2}$$

$$\times \frac{|E_B|}{(a_0^d W^2)^2}\exp\left(-\frac{I_4}{16\pi} - \frac{I_4|E_B|^{1/2}}{2a_0^d W^2}\right), \quad d = 3,$$

$$\frac{v(E_B)}{v_0(-E_B)} = \frac{I_6 - I_4}{8\pi^2}\left|I_4 \frac{\bar{D}_R(1/3)}{\bar{D}_R(1)}\right|^{1/2}\left(\frac{4\pi|E_B|}{a_0^d W^2}\right)^{3/2 - I_4/8\pi}$$

$$\times \exp\left(-\frac{I_4}{8\pi} - \frac{I_4|E_B|}{2a_0^d W^2}\right), \quad d = 2, \quad (5.59)$$

where the numerical values of the parameters are† [62]:

$$\bar{D}_R(1) \approx 10.544, \quad \bar{D}_R\left(\frac{1}{3}\right) \approx 1.4571, \quad I_4 \approx 75.589,$$

$$I_6 \approx 659.87, \quad d = 3,$$

$$\bar{D}_R(1) \approx 135.3, \quad \bar{D}_R\left(\frac{1}{3}\right) \approx 1.465, \quad I_4 \approx 23.402,$$

$$I_6 \approx 71.080, \quad d = 2.$$

At $d < 2$, there are no divergences in the determinants, and result (5.58) holds in terms of the bare values [that is, with the replacement $E \to E_B$, $\bar{D}_R(1) \to \bar{D}(1)$, $\bar{D}_R(1/3) \to \bar{D}(1/3)$]. At $d = 1$, the instanton has the form $\phi_c(x) = \sqrt{2}/\cosh x$, and the eigenvalues of Eqn (5.36) are $\mu_s = s(s+1)$, $s = 1, 2, \ldots$ [39]. Calculation of the parameters in Eqn (5.58)

$$\bar{D}(1) = -\frac{1}{5}, \quad \bar{D}\left(\frac{1}{3}\right) = \frac{1}{3}, \quad I_4 = \frac{16}{3}, \quad I_6 = \frac{128}{15} \quad (5.60)$$

gives

$$v(E_B) = \frac{4}{\pi}\frac{|E_B|}{a_0^d W^2}\exp\left\{-\frac{8|E_B|^{3/2}}{3a_0^d W^2}\right\}, \quad (5.61)$$

which coincides with the exact result reported by Gal'perin [40, 63].

---

† The final expressions (16) in Ref. [43] contain obvious misprints, but the numerical coefficients agree with Eqn (5.59).

## 6. $(4 - \epsilon)$-Dimensional theory

In this section it will be convenient to denote the coefficient at $|\varphi|^4$ in Hamiltonian (1.14) by $u$,

$$H\{\varphi\} = \int d^d x \left\{ \frac{1}{2} |\nabla \varphi|^2 + \frac{1}{2} \varkappa_0^2 |\varphi|^2 + \frac{1}{4} u |\varphi|^4 \right\}, \qquad (6.1)$$

retaining the notation $g_0$ for the dimensionless quantity

$$g_0 = u \Lambda^{-\epsilon}. \qquad (6.2)$$

### 6.1 Structure of the approximation for $\Sigma(0, \varkappa)$

At $d = 4 - \epsilon$, the perturbation-theory expansion for the self-energy $\Sigma(p, \varkappa)$ with $p = 0$ has the structure

$$\varkappa^2 + \Sigma(0, \varkappa) - \Sigma(0, 0) \equiv \varkappa^2 Y(\varkappa)$$
$$= \varkappa^2 \sum_{N=0}^{\infty} (u\Lambda^{-\epsilon})^N \sum_{K=0}^{N} A_N^K(\epsilon) \left[ \frac{(\Lambda/\varkappa)^\epsilon - 1}{\epsilon} \right]^K, \qquad (6.3)$$

where the coefficients $A_N^K(\epsilon)$ are finite at $\epsilon \to 0$, and $A_0^0(\epsilon) \equiv 1$. Expansion (6.3) follows from the fact that the quantity $Y$ in the $N$th order of the expansion theory is a homogeneous polynomial of order $N$ composed of $\Lambda^{-\epsilon}$ and $\varkappa^{-\epsilon}$: indeed, in going from a diagram of order $N$ to a diagram of order $(N + 1)$, the dimensionality with respect to momentum is reduced by $\epsilon$ [46], which gives rise to a factor of $\Lambda^{-\epsilon}$ or $\varkappa^{-\epsilon}$, depending on whether the corresponding contribution is determined by large or small momenta. Factoring out the multipliers $\epsilon^{-K}$ from the coefficients $A_N^K(\epsilon)$ provides the correspondence with Eqn (4.11) at $\epsilon \to 0$.

After expansion of the coefficients $A_N^K(\epsilon)$ in $\epsilon$,

$$A_N^K(\epsilon) = \sum_{L=0}^{\infty} A_N^{K,L} \epsilon^L, \qquad (6.4)$$

and passage to the limit $\Lambda \to 0$, expansion (6.3) becomes similar to Eqn (4.11), where the large logarithms are replaced by powers of $1/\epsilon$. The standard procedure of $\epsilon$ expansion [9, 10] retains a few senior powers of $1/\epsilon$ in each order of the perturbation theory; the first $\epsilon$ approximation is confined to the coefficients $A_N^{N,0}$ that coincide with the coefficients at the main logarithms in Eqn (4.11). As in the case of $d = 4$, this approximation is not sufficient when $u < 0$ because of the faster growth of the coefficients at the lower powers of $1/\epsilon$ with increasing $N$. It is only possible to restrict the expansion to the coefficients $A_N^{N,0}$ when $N \sim 1$, whereas for large $N$ one has to take all $A_N^{K,L}$ into account, calculating them in the main asymptotics with respect to $N$.

According to Eqn (6.3), the quantity $Y$ is a function of $g_0$ and $\Lambda/\varkappa$; it is multiplicatively renormalizable and satisfies the Callan–Symanzik equation [38, 39]:

$$\left[ \frac{\partial}{\partial \ln \Lambda} + W(g_0, \epsilon) \frac{\partial}{\partial g_0} + V(g_0, \epsilon) \right] Y = 0. \qquad (6.5)$$

Functions $W(g_0, \epsilon)$ and $V(g_0, \epsilon)$ are expanded in a series

$$W(g_0, \epsilon) = \sum_{M=1}^{\infty} W_M(\epsilon) g_0^M = \sum_{M=1}^{\infty} \sum_{M'=0}^{\infty} W_{M,M'} g_0^M \epsilon^{M'},$$

$$V(g_0, \epsilon) = \sum_{M=1}^{\infty} V_M(\epsilon) g_0^M = \sum_{M=1}^{\infty} \sum_{M'=0}^{\infty} V_{M,M'} g_0^M \epsilon^{M'}, \qquad (6.6)$$

whose first coefficients were calculated in Ref. [46]†:

$$W_1(\epsilon) = -\epsilon, \qquad W_{2,0} = K_4(n + 8),$$
$$W_{3,0} = -3K_4^2(3n + 14), \qquad V_{1,0} = -K_4(n + 2). \qquad (6.7)$$

Substituting Eqn (6.3) and (6.6) into (6.5), we obtain a set of equations in the coefficients $A_N^K(\epsilon)$

$$(K + 1) A_N^{K+1}(\epsilon) = (N - K) \epsilon A_N^K(\epsilon)$$
$$- \sum_{M=1}^{N-K} \left[ (N - M) W_{M+1}(\epsilon) + V_M(\epsilon) \right] A_{N-M}^K(\epsilon), \qquad (6.8)$$

from which one can easily derive a similar set of equations in $A_N^{K,L}$ in expansion (6.4). Wilson's method [9, 10] is based on the fact that in the $n$th $\epsilon$ approximation one needs to know the coefficients $A_N^{N-K,L}$ with $K + L \leqslant n - 1$, for which there is a closed system of difference equations

$$-N x_N = \left[ W_{2,0}(N - 1) + V_{1,0} \right] x_{N-1},$$

$$-(N - 1) y_N = \left[ W_{2,0}(N - 1) + V_{1,0} \right] y_{N-1}$$
$$+ \left[ W_{3,0}(N - 2) + V_{2,0} \right] x_{N-2},$$

$$-N z_N = \left[ W_{2,0}(N - 1) + V_{1,0} \right] z_{N-1}$$
$$+ \left[ W_{2,1}(N - 1) + V_{1,1} \right] x_{N-1} - y_N, \qquad (6.9)$$
$$\ldots\ldots,$$

which is solved by the method of variation of parameters [64] and allows consecutively $x_N \equiv A_N^{N,0}$, $y_N \equiv A_N^{N-1,0}$, $z_N \equiv A_N^{N,1}, \ldots$ to be found. In order to define the initial conditions for this system and to find $W_{2,0}, V_{1,0} \ldots$, one has to calculate a few first orders of the perturbation theory. Separating the main asymptotics in $N$, one can easily prove by induction that

$$A_N^{N-K,L} = C_{K+L}^K (-W_{2,0})^N \frac{\Gamma(N - \beta_0)}{\Gamma(N + 1)\Gamma(-\beta_0)}$$
$$\times \frac{(-W_{3,0})^{K+L}}{(-W_{2,0})^{2(K+L)}} \frac{(N \ln N)^{K+L}}{(K + L)!}, \qquad (6.10)$$

where $\beta_0 = -V_{1,0}/W_{2,0} = (n + 2)/(n + 8)$. For the parquet coefficients $A_N^{N,0}$, the result (6.10) is exact.

Wilson's method turns out not to be efficient for studying higher orders in $\epsilon$, and it is more convenient to start with the set of equations (6.8), which structurally is a recursion relation for calculating $A_N^{K+1}(\epsilon)$ with the given $A_N^K(\epsilon)$, $A_{N-1}^K(\epsilon), \ldots, A_K^K(\epsilon)$. Information about the coefficients $A_N^K(\epsilon)$ with $N \gg 1$ can be obtained using Lipatov's method; according to Section 5.5, the $N$th coefficient of expansion of $\Sigma(p, \varkappa)$ in powers of $u$ is given by Eqn (5.51). Representing Eqn (5.51) in the form of expansion (6.3), we get

$$A_N^K(\epsilon) = \tilde{c}_2 \Gamma(N + b) a^N C_N^K \int_0^\infty d \ln R^2$$
$$\times R^{-2} \left\{ \epsilon + \frac{2K_d I_4}{N} \exp[f(R) - \epsilon \ln R] \right\}^K$$
$$\times \exp \left[ -N f(R) + N \epsilon \ln R + 2K_d I_4(R) \frac{1 - R^{-\epsilon}}{\epsilon} \right], \qquad (6.11)$$

where $\tilde{c}_2 = c_2 \langle \phi_c^3 \rangle_0^2$.

---

† According to Refs [38, 39], the function $V(g_0, \epsilon)$ coincides with $\eta_2(g_0, \epsilon)$ introduced in Ref. [46]. Note that one has to distinguish the functions $W$ and $\eta_2$ for different renormalizations (see Appendix 1 in Ref. [38]).

Lipatov's method accurately reconstructs the coefficients $A_N^K(\epsilon)$ only for $K \ll N$, which is explained by their fast decrease with increasing $K$ and by the limited accuracy ($\sim 1/N$) of the main asymptotics. It is easy to see that at $K \ll N$ equation (6.8) is satisfied by the result (6.11)† if we only retain the term with $M=1$ in the sum with respect to $M$. The latter is made possible by the factorial growth of $A_N^K(\epsilon)$ with respect to $N$ under the assumption that $W_N(\epsilon)$ and $V_N(\epsilon)$ (which occur in the equations with $K=0,1$) grow more slowly than $A_N^0(\epsilon)$. This may be regarded as a consequence of the validity of Eqn (6.11) for $K=0,1,2$.

The set of equations (6.8) defines $A_N^K(\epsilon)$ with $K > 0$ for the given $A_N^0(\epsilon)$. Since Eqn (6.11) holds for the latter for all $N \gg 1$, it can be used as the boundary condition for Eqn (6.8), which allows all $A_N^K(\epsilon)$ with large $N$ to be found. Then, retaining for $N \sim 1$ only the leading order in $1/\epsilon$ as defined by the coefficients (6.11), one can easily find the sum of the series (6.3).

The recurrence relation (6.8) can be reduced for large $N$ owing to the factorial growth of $A_N^K(\epsilon)$, so that $A_{N-1}^K(\epsilon) \sim A_N^K(\epsilon)/N$. As a matter of fact, on the right-hand side of Eqn (6.8) it is only necessary to retain the terms with $A_N^K(\epsilon), A_{N-1}^K(\epsilon), A_{N-2}^K(\epsilon)$. The fact is that the formally senior term in $N$ with $A_N^K(\epsilon)$ contains a small factor $\epsilon$, and for $N\epsilon \lesssim 1$ is small compared to $A_{N-1}^K(\epsilon)$, whereas the necessarily small term with $A_{N-2}^K(\epsilon)$ is required for achieving the desired accuracy: the calculation of an arbitrary $A_N^K(\epsilon)$ given the known $A_N^0(\epsilon)$ takes $\sim N$ iterations, which leads to an accumulation of error if the accuracy of each iteration is $\sim 1/N$. The reduced equation can be formally solved by direct iteration, whereupon the result can be simplified in the limit of large $N$.

By analogy with the case of $d=4$ (see Fig. 7), we may mark region $I$ in the $(N,K)$ plane (Fig. 9), in which the coefficients $A_N^K(\epsilon)$ are determined by the coefficients $A_N^0(\epsilon)$ with large $N$, for which the Lipatov asymptotics hold, and region $II$ ($M \ll \ln N$, $N\epsilon \ll 1$), 'governed' by the trivial coefficient $A_0^0(\epsilon) = 1$. In between, there is the nonuniversality region $III$ ($M \sim \ln N$, $N\epsilon \lesssim 1$), which requires knowing the coefficients $A_N^0(\epsilon)$ with $N \sim 1$; the latter region does not give any significant contribution to the sum (6.3).

Let us quote the results for the coefficients $A_N^N(\epsilon)$, which determine the sum (6.3) in the continuum limit $\Lambda \to \infty$. In the ranges $Nt > 1$ or $1 - Nt \ll \epsilon^{1/2}$ ($t$ is defined below), we have

$$A_N^N(\epsilon) = \tilde{c}_2 \Gamma(N-\beta)\epsilon^N a^N F(N), \quad (6.12)$$

where

$$F(N) = \left(\frac{t}{2\pi}\right)^{1/2} \exp\left[f_\infty(Nt\ln N - 1) + \frac{1}{t}\right]$$
$$\times \int_0^\infty \mathrm{d}x \exp\left[-\frac{t}{2}\left(N - \frac{1}{t} - x\right)^2\right] x^{b+\beta-f_\infty Nt} J(x), \quad (6.13)$$

$$J(N) = \int_0^\infty \mathrm{d}\ln R^2 R^{2/3} \exp\left[-Nf(R) + N\epsilon \ln R\right], \quad (6.14)$$

$$t = -\frac{\epsilon a}{W_2(\epsilon)} \xrightarrow{\epsilon \to 0} \frac{3\epsilon}{n+8}, \quad \beta(\epsilon) = -\frac{V_1(\epsilon)}{W_2(\epsilon)} \xrightarrow{\epsilon \to 0} \beta_0,$$

$$f_\infty = \frac{W_3(\epsilon)}{aW_2(\epsilon)} \xrightarrow{\epsilon \to 0} \frac{3n+14}{n+8}. \quad (6.15)$$

For $N\epsilon \ll 1$, we have a result of the form (6.4) with the coefficients $A_N^{K,L}$ defined by Eqn (6.10).

## 6.2 Renormalization of the energy and damping

The two important contributions to the sum (6.3), the nonperturbative and quasi-parquet, come, respectively, from regions $I$ and $II$ (see Fig. 9). The quasi-parquet contribution to the sum (6.3) is determined by the coefficients $A_N^{N-K,L}$ with $K, L \sim 1$ and is calculated on the basis of Eqn (6.10). At $\Lambda \to \infty$, it can be written in the form

$$[Y(\varkappa)]_{\text{quasiparq}} = \left(1 + \frac{W_{2,0}\tilde{u}\varkappa^{-\epsilon}}{\epsilon}\right)^{\beta_0},$$

$$\tilde{u} \approx u\left(1 + \frac{W_{3,0}}{W_{2,0}^2}\epsilon \ln \epsilon\right), \quad (6.16)$$

which differs from the parquet result [52] only in that $u$ is replaced by $\tilde{u}$.

For calculating the nonperturbative contribution we use Eqn (6.12) and sum up Eqn (6.3) with respect to $N$ from a certain large $N_0$ to infinity:

$$[\Sigma(0,\varkappa)]_{\text{nonpert}} \equiv \mathrm{i}\Gamma_0(\varkappa^2)$$
$$= \mathrm{i}\pi\tilde{c}_2\varkappa^2 \left(\frac{\varkappa^\epsilon}{au}\right)^b \exp\left(-\frac{\varkappa^\epsilon}{au}\right) F\left(\frac{\varkappa^\epsilon}{au}\right). \quad (6.17)$$

Here we have used the formula

$$\mathrm{Im} \sum_{N=N_0}^\infty \Gamma(N+b) a^N (g_0 - \mathrm{i}0)^N f(N)$$
$$= \frac{\pi}{(ag_0)^b} \exp\left(-\frac{1}{ag_0}\right) f\left(\frac{1}{ag_0}\right), \quad ag_0 > 0, \quad (6.18)$$

which holds for slowly changing functions $f(N)$; it is derived by expanding $f(N)$ into the Fourier integral, making use of

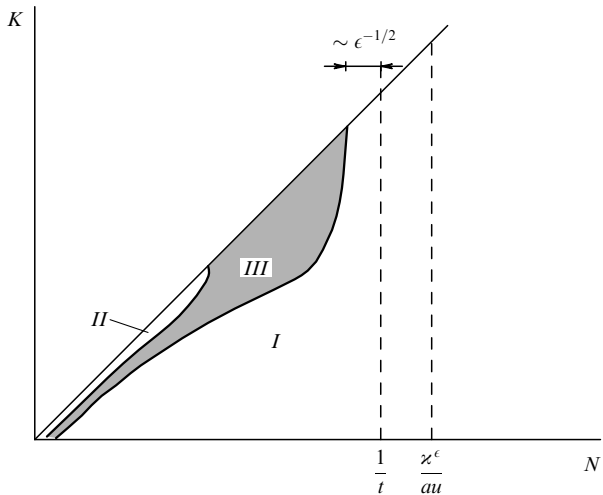

**Figure 9.** $I$ and $II$, regions which give the nonperturbative and quasi-parquet contributions into (6.3), respectively; $III$, the region of nonuniversality. The parameter $t \sim \epsilon$ is defined in Eqn (6.15). The nonperturbative contribution is effectively evaluated at $N = \varkappa^\epsilon/au$; the inequality $\varkappa^\epsilon/au > 1/t$ arises as a result of the solution of Eqn (6.19).

† Here we have in mind the condition $N \gg 1$. Under the stronger condition $N\epsilon \gg 1$, Lipatov's method gives the coefficients for all $K$, which can be easily proved in the same way.

Eqn (4.10), and retaining only the long-wave Fourier components. The arbitrariness in the definition of $f(N)$ at $N < N_0$ permits the condition of slow change to be satisfied by any function $f(N)$ that at large $N$ does not change faster than a power-law function. The unusual phenomenon associated with the divergence of the series is that the sum in Eqn (6.18) is determined by arbitrarily large $N$ (and therefore the result does not depend on $N_0$); the result, however, involves the value of $f(N)$ at a finite $N = 1/ag_0$. Because of this, the correction factor which distinguishes Eqn (6.12) from (6.11) and equals unity as $N \to \infty$, is effectively evaluated at $N = \varkappa^\epsilon/au$ and is important.

Including the contributions (6.16) and (6.17) in the sum (6.3), we obtain the equation

$$\varkappa_0^2 - \varkappa_c^2 = \varkappa^2 \left(1 + \frac{8K_4 \tilde{u} \varkappa^{-\epsilon}}{\epsilon}\right)^{1/4} + i\Gamma_0(\varkappa^2), \quad \varkappa^2 = -E - i\Gamma, \tag{6.19}$$

where $\varkappa_c^2 = \Sigma(0,0)$, and $\varkappa_0^2 = \varkappa^2 + \Sigma(0,\varkappa)$. Assuming that

$$\varkappa^2 = |\varkappa|^2 \exp(-i\varphi), \quad x = \frac{2}{\epsilon}\left[\left(\frac{|\varkappa|^2}{\Gamma_c}\right)^{\epsilon/2} - 1\right],$$

$$\Gamma_c = \left(\frac{8K_4|\tilde{u}|}{\epsilon}\right)^{2/\epsilon}, \tag{6.20}$$

separating the real and imaginary parts in Eqn (6.19), and taking into account that the solution only exists for large $x$, we get the linkage of the damping $\Gamma$ and the renormalized energy $E$ with the bare energy $E_B = -\varkappa_0^2$ in parametric form

$$\Gamma = \Gamma_c \left(1 + \frac{\epsilon x}{2}\right)^{2/\epsilon} \sin\varphi, \quad E = -\Gamma_c \left(1 + \frac{\epsilon x}{2}\right)^{2/\epsilon} \cos\varphi, \tag{6.21}$$

$$-E_B + E_c = \Gamma_c \left(1 + \frac{\epsilon x}{2}\right)^{2/\epsilon} \left(\frac{\epsilon x/2}{1 + \epsilon x/2}\right)^{1/4}$$

$$\times \left[\cos\left(\varphi + \frac{\varphi}{4x}\right) - \tan\frac{\varphi(1+2\epsilon x)}{3}\sin\left(\varphi + \frac{\varphi}{4x}\right)\right]. \tag{6.22}$$

Here $x(\varphi)$ is a single-valued function on the interval $0 < \varphi < \pi$ as shown in Fig. 10, which satisfies the equation

$$\sin\left(\varphi + \frac{\varphi}{4x}\right) = \frac{\exp(-4x/3)}{x^{1/4}} I(x) \cos\frac{\varphi(1+2\epsilon x)}{3}. \tag{6.23}$$

Function $I(x)$ is given by

$$I(x) = \tilde{c}_2 \left(\frac{3}{4}\right)^{1/4} \left(\frac{\pi t}{2}\right)^{1/2}$$

$$\times \exp\left[-f_\infty + f_\infty \frac{2+\epsilon x}{2} \ln\frac{\bar{A}(1+\epsilon x/2)}{t}\right]$$

$$\times \int_0^\infty dz \exp\left[-\frac{t}{2}\left(\frac{\epsilon x}{2t} - z\right)^2\right] z^{b+\beta-f_\infty(1+\epsilon x/2)} J(z), \tag{6.24}$$

where

$$\bar{A} \approx \frac{7}{8} \epsilon \ln\frac{1}{\epsilon},$$

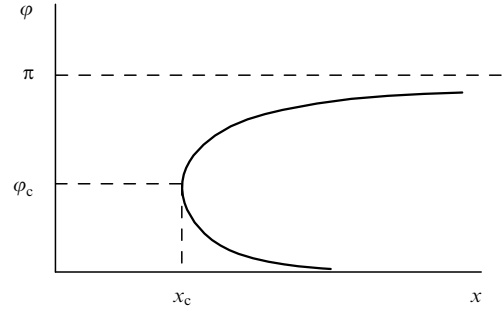

**Figure 10.** Form of the function $\varphi(x)$.

and the quantity $E_c$ defines the shift of the edge of the band and in the one-loop approximation is

$$E_c = -W^2 \int \frac{d^d p}{(2\pi)^d} \frac{1}{\epsilon(p)}. \tag{6.25}$$

Equations (6.21), (6.23) are much simplified in the two overlapping regions. In the region of large $|E|$, $x \gg \ln(1/\epsilon)$, when the right-hand side of Eqn (6.23) is small and $\varphi$ is close to 0 or $\pi$, the function $\Gamma(E)$ is asymptotically approximated by

$$\Gamma(E) = \begin{cases} \left(\frac{\pi}{8}\right)\epsilon E\left[\left(\frac{E}{\Gamma_c}\right)^{\epsilon/2} - 1\right]^{-1}, & E \gg \Gamma, \\ \Gamma_0(E)\left[1 - \left(\frac{|E|}{\Gamma_c}\right)^{-\epsilon/2}\right]^{-1/4}, & -E \gg \Gamma, \end{cases} \tag{6.26}$$

which creates the illusion of a spurious pole [20, 27] $[\Gamma_0(E) \equiv \Gamma_0(|\varkappa|^2)]$; for large positive $E$, the result of the kinetic equation is reconstructed, while for large negative $E$, the damping becomes purely nonperturbative. In the neighborhood of the spurious pole, $x \lesssim \epsilon^{-1/2}$, the function $I(x)$ can be replaced by $I(0) \sim \epsilon^{-7/12}[\ln(1/\epsilon)]^{17/12}$. The minimum value of $x$ with logarithmic accuracy is

$$x_{\min} \approx \frac{7}{16} \ln\frac{1}{\epsilon}, \tag{6.27}$$

so that the spurious pole is passed around at a distance of the order of $\epsilon \ln(1/\epsilon)$, and the effective interaction (see Section 3) remains logarithmically weak†.

### 6.3 Self-energy $\Sigma(p, \varkappa)$ for finite momenta

Calculation of the density of states requires knowing the self-energy $\Sigma(p, \varkappa)$ for finite momenta. As in the case of $p = 0$, this quantity is comprised of the nonperturbative and quasi-parquet contributions; let us consider the calculation of these.

**6.3.1 Parquet approximation.** First of all, let us calculate $\Sigma(p, \varkappa)$ in the parquet approximation, which can be done through a certain extension of Ginzburg's paper [52]. From Ward's identity we obtain

$$\frac{\partial G^{-1}(p, \varkappa)}{\partial \varkappa_0^2} \delta_{\alpha\beta} = \delta_{\alpha\beta} - \frac{1}{2} \sum_\sigma \int \frac{d^d q}{(2\pi)^d} G(q)^2 \Gamma_{\alpha\sigma\sigma\beta}^{(0,4)}(p, -q, q, -p), \tag{6.28}$$

---

† In the limit $\epsilon \to 0$, one must bear in mind that the cutoff parameter $\Lambda$ should be finite, because of which the distance to the spurious pole remains finite.

where $\Gamma_{\alpha\beta\mu\nu}^{(0,4)}(p_1,p_2,p_3,p_4)$ is a complete vertex with four legs $(p_1+p_2+p_3+p_4=0)$. As is common in the parquet calculations [52, 55], at the vertex we go over from the momenta $p_1$, $p_2, p_3, p_4$ to $p'$, $p$, $q$, where

$$2p' = p_4 - p_3, \quad 2p = p_1 - p_2, \quad q = p_1 + p_2 = -p_3 - p_4, \quad (6.29)$$

and the legs of the vertex are numbered in such a way that $p' > p > q > 0$.

In the main logarithmic approximation at $d=4$, the calculation of $\Gamma^{(0,4)}(p',p,q)$ requires summing up the parquet diagrams (Fig. 11a) obtained by consecutive splitting of simple vertexes into two parts connected with two lines. As the order of a diagram is increased by one, the smallness of $\sim g_0$, associated with the additional vertex, is compensated by the large logarithm, associated with the additional pair of lines [11]. For $d=4-\epsilon$, in place of the large logarithm we have the parameter $1/\epsilon$. Using the results of Ref. [52] for $\Gamma^{(0,4)}$ in Eqn (6.28), and taking into account the linkage between $G^{-1}(p,\varkappa)$ and $\Sigma(p,\varkappa)$, we get the desired result for the parquet approximation:

$$\Sigma(p,\varkappa) = \Sigma(0,0) + \varkappa^2 \left\{ t(x_\infty)^{(n+2)/(n+8)} \right.$$

$$\left. + \frac{6}{n-4}\left[\frac{t(x_\infty)}{t(x)}\right]^{(n+2)/(n+8)} - \frac{n+2}{n-4}\left[\frac{t(x_\infty)}{t(x)}\right]^{6/(n+8)} \right\},$$

$$t(x) = 1 + K_d \Lambda^{-\epsilon}(n+8)ux, \quad x = \frac{(\Lambda/p)^\epsilon - 1}{\epsilon},$$

$$x_\infty = \frac{(\Lambda/\varkappa)^\epsilon - 1}{\epsilon}, \quad (6.30)$$

whence the result for the disordered system follows as $n \to 0$. Formula (6.30) holds for $p \gtrsim \varkappa$, and therefore the limit $p \to 0$ is interpreted as $p \to \varkappa$.

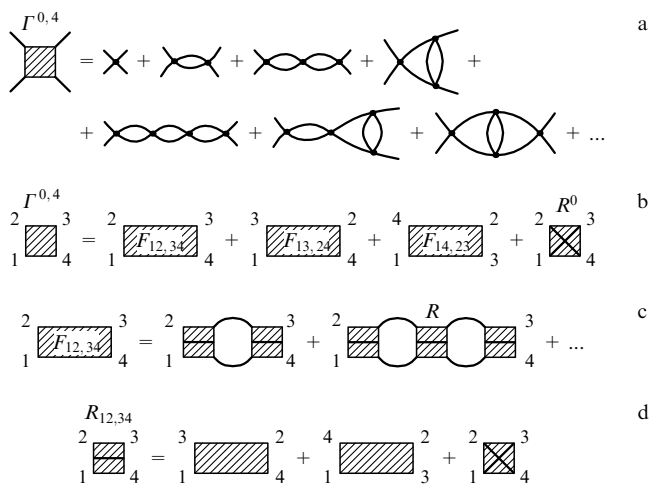

**Figure 11.** (a) Parquet series of diagrams for $\Gamma^{(0,4)}$ (obtained by consecutive splitting of simple vertexes into two parts connected with a pair of lines); (b–d) set of equations for the complete vertex $\Gamma^{(0,4)}$, three 'blocks' $F_{ij,kl}$, three 'crossed' vertexes $R_{ij,kl}$, and the irreducible vertex $R^0$. The approximation $R^0 = -2u$ corresponds to the summation of the sequence of diagrams (a).

**6.3.2 Parquet in higher orders in $\epsilon$.** At $p=0$, the quasi-parquet contribution to $\Sigma(0,\varkappa)$ is determined by the coefficients $A_N^{N-K,L}$ with $K,L \sim 1$, and the coefficients with $K,L \neq 0$ are only important when $N$ is large. Now we have the problem of formulating a similar quasi-parquet approximation for finite momenta. We need to calculate $\Sigma(p,\varkappa)$ in the arbitrary finite order in $\epsilon$, but only taking into account the principal asymptotics in $N$ for the nonparquet coefficients of the expansion.

As we saw earlier, the calculation of $\Sigma(p,\varkappa)$ requires knowing the four-leg vertex $\Gamma^{(0,4)}(p,k,q)$ that depends on three significantly different momenta, $p \gg k \gg q \gg \varkappa$. The method used above allows us to find the quasi-parquet approximation to $\Gamma^{(0,4)}$ at $p \sim k \sim q \gg \varkappa$: writing down an expansion similar to Eqn (6.3)

$$\Gamma^{(0,4)}(p,p,p) = \sum_{N=1}^{\infty}(u\Lambda^{-\epsilon})^N \sum_{K=0}^{N-1} A_N^K(\epsilon)\left[\frac{(\Lambda/p)^\epsilon - 1}{\epsilon}\right]^K \quad (6.31)$$

[with another coefficients $A_N^K(\epsilon)$], and taking advantage of the fact that the vertex $\Gamma^{(0,4)}$ is multiplicatively renormalizable [46] and satisfies the Callan–Symanzik equation (4.20)†, we obtain an equation similar to Eqn (6.8), from which we may get a result of the type of Eqn (6.10), whereupon the summation over region $II$ yields

$$\left[\Gamma^{(0,4)}(p,p,p)\right]_{\text{quasiparq}}$$

$$= \frac{-2u}{\Delta + (W_3/W_2)up^{-\epsilon}\ln\Delta}\bigg|_{\Delta=1+W_2up^{-\epsilon}/\epsilon} \approx \frac{-2\tilde{u}}{1 + W_2\tilde{u}p^{-\epsilon}/\epsilon}. \quad (6.32)$$

The logarithmic term is only important near the root of the quantity $\Delta$, and one can make the replacements $up^{-\epsilon} \to -\epsilon/W_2$ and $\Delta \to \bar{\Delta}$.

It is known [11, 53] that the parquet approximation can be obtained from the standpoint of a general structural analysis of diagrams: the complete vertex $\Gamma^{(0,4)}$ is represented as three 'blocks' $F_{ij,kl}$ and the irreducible four-leg vertex $R^0$ (Fig. 11b); each block is a sum of the diagrams reducible with respect to the pair of lines in the corresponding channel and derived by repeating the 'crossed' vertex $R$ (Fig. 11c), which in turn is the sum of two blocks and an irreducible four-leg vertex (Fig. 11d). Constructing equations of the type (c), (d) for the other two blocks, we get a set of 7 equations in 8 variables [53]; all variables are uniquely defined by fixing the vertex $R^0$, so that

$$\Gamma^{(0,4)}(p,k,q) = F\{R^0(p,k,q)\}, \quad (6.33)$$

where $F\{\ldots\}$ is some functional. The parquet approximation corresponds to replacing the irreducible four-leg vertex by a simple vertex, $R^0(p,k,q) = -2u$, since the system (b–d) then corresponds to the summation of diagrams (a) (see Fig. 11).

In Appendix 2 of Ref. [38] it is proved that for $d=4$ the vertex $R^0$ only depends on the maximum momentum; by virtue of continuity, this property applies also to small $\epsilon$. Given this, Eqn (6.33) becomes

$$\Gamma^{(0,4)}(p,k,q) = F\{R^0(p,p,p)\}. \quad (6.34)$$

---

† In which $V(g_0,\epsilon) \equiv -2\eta(g_0,\epsilon)$, where $\eta(g_0,\epsilon)$ is the function introduced in Ref. [46].

Assuming that $k = q = p$, and inverting Eqn (6.34), we can derive the approximation for $R^0$ corresponding to the result (6.32), after which Eqn (6.34) in principle gives a solution to the problem.

The functional $F\{\ldots\}$ is defined by the set of equations (b–d) (Fig. 11), which is extremely complicated and does not seem to have ever been solved. Usually one resorts to the simplified parquet equations derived by the method developed by Sudakov [54] and modified by Polyakov [55]. This method does not assume any particular approximation for $R^0$, but is based on the logarithmic calculation of the integrals, in which the limits of integration are only evaluated by order of magnitude. A similar technique can also be applied for the case of $d = 4 - \epsilon$, for example:

$$\int_0^\Lambda \frac{k^{d-1}\,\mathrm{d}k}{(k^2+\varkappa^2)^2} = \int_{\sim\varkappa}^\Lambda \frac{k^{d-1}\,\mathrm{d}k}{k^4} = \frac{\varkappa^{-\epsilon} - \Lambda^{-\epsilon}}{\epsilon} + O(\epsilon^0). \quad (6.35)$$

The lower limit in the second integral is only known by order of magnitude, but this uncertainty has no effect on the main contribution since the replacement $\varkappa \to c\varkappa$ gives rise to a factor of $c^{-\epsilon} \approx 1 - \epsilon \ln c$ and only affects the quantity $O(\epsilon^0)$. At first sight, this approach is only justified in the leading order in $\epsilon$; indeed, the replacement $\varkappa \to c\varkappa$ in expression (6.3) with $\Lambda \to \infty$ in the terms of the leading order in $\epsilon$ affects the coefficients at the higher powers of $\epsilon$,

$$A_N^{N,L} \to A_N^{N,L} + \frac{(-\ln c)^L}{L!} N^L A_N^{N,0}, \quad L \sim 1. \quad (6.36)$$

According to Eqn (6.10), however, we have $A_N^{N,L} \sim A_N^{N,0}(N\ln N)^L$, and the principal asymptotics with respect to $N$ do not sense the replacement $\varkappa \to c\varkappa$. Because of this, such an approach is acceptable for arbitrary finite order in $N$ when only the leading order with respect to $\epsilon$ is considered. This allows us to apply the entire parquet scheme to the calculation of the quasi-parquet contribution to $\Gamma^{(0,4)}$.

Substituting Eqn (6.32) into the parquet equation [52, 53, 55]

$$\Gamma^{(0,4)}(x,x,x) = R^0(x,x,x)$$
$$+ \frac{1}{2} K_d \Lambda^{-\epsilon}(n+8)\int_0^x \mathrm{d}t\,\left[\Gamma^{(0,4)}(t,t,t)\right]^2, \quad (6.37)$$

we get $R^0(x,x,x) = -2\tilde{u}$, and the sum $\left[\Sigma(p,\varkappa)\right]_{\text{quasiparq}}$ is obtained from the parquet result (6.30) by replacing $u$ with $\tilde{u}$.

**6.3.3 Nonperturbative contribution.** The nonperturbative contribution (see Section 6.4) is only important in the range of large negative $E$ and can be calculated directly from Lipatov's asymptotics (5.51) (at $N = \varkappa^\epsilon/au \gg 1/\epsilon$, the correction factor which distinguishes the results of the type of Eqn (6.12) and (6.11) is equal to unity):

$$\left[\Sigma(p,\varkappa)\right]_{\text{nonpert}} = \mathrm{i}\pi c_2\varkappa^2\left(\frac{\varkappa^\epsilon}{au}\right)^b \exp\left(-\frac{\varkappa^\epsilon}{au}\right)$$
$$\times \int_0^\infty \mathrm{d}\ln R\, R^{-2}\langle\phi_c^3\rangle_{pR/\varkappa}\langle\phi_c^3\rangle_{-pR/\varkappa}$$
$$\times \exp\left\{-\frac{\varkappa^\epsilon}{au}\left[f(R) - \epsilon\ln R\right] + \frac{2K_d I_4(R)}{\epsilon}\right\}. \quad (6.38)$$

At $p = 0$, the integral is determined by the neighborhood of the saddle point $R_0$, which is the root of the equation

$$\epsilon = 6R_0^2(-\ln R_0 + \ln 2 - C - 1), \quad (6.39)$$

so that $R_0 \approx \left[\epsilon/3\ln(1/\epsilon)\right]^{1/2}$. At $p \lesssim \varkappa R_0^{-1}$, expression (6.38) does not depend on $p$; at $p \gtrsim \varkappa R_0^{-1}$, it falls off rapidly with increasing $p$. Since the subsequent calculations are of the logarithmic accuracy (see Section 6.4), we can settle for the result

$$\left[\Sigma(p,\varkappa)\right]_{\text{nonpert}} \approx \left[\Sigma(0,\varkappa)\right]_{\text{nonpert}}\theta(\varkappa R_0^{-1} - p). \quad (6.40)$$

The final result for $\Sigma(p,\varkappa)$ is given by the sum of the quasi-parquet and nonperturbative contributions

$$\Sigma(p,\varkappa) - \Sigma(0,\varkappa) = \varkappa^2\left\{1 - \frac{3}{2}\left[\frac{t(x)}{t(x_\infty)}\right]^{-1/4} + \frac{1}{2}\left[\frac{t(x)}{t(x_\infty)}\right]^{-3/4}\right\}$$
$$- \mathrm{i}\Gamma_0(\varkappa^2)\theta(p - \varkappa R_0^{-1}), \quad (6.41)$$

where

$$t(x) = 1 + \frac{8K_4\tilde{u}x}{\epsilon}, \quad x = p^{-\epsilon}, \quad x_\infty = \varkappa^{-\epsilon}. \quad (6.42)$$

### 6.4 Calculation of the density of states

The density of states $v(E)$ is given by

$$v(E) = \frac{1}{\pi}\operatorname{Im}\int\frac{\mathrm{d}^d p}{(2\pi)^d}\,G(p,\varkappa) \equiv \frac{1}{\pi}\operatorname{Im}Y(\varkappa)\bigg|_{\varkappa^2=-E-\mathrm{i}\Gamma}. \quad (6.43)$$

The main contribution to the integral comes from the region $|p| \gtrsim \varkappa$, for which

$$Y(\varkappa) \approx \int\frac{\mathrm{d}^d p}{(2\pi)^d}\frac{1}{p^2} - \int\frac{\mathrm{d}^d p}{(2\pi)^d}\frac{\varkappa^2 - \Sigma(p,\varkappa) + \Sigma(0,\varkappa)}{p^4}. \quad (6.44)$$

Substituting Eqn (6.41) into (6.44) and eliminating $\Gamma_0(\varkappa^2)$ with the aid of Eqn (6.19), we obtain

$$v = \frac{\Gamma_c}{4\pi|\tilde{u}|}\left(1 + \frac{\epsilon x}{2}\right)^{2/\epsilon}\left[\left(1 + \frac{2}{\epsilon x}\right)^{-1/4}\left(1 - \frac{R_0^\epsilon}{2+\epsilon x}\right)\right.$$
$$\times \sin\left(\varphi + \frac{\varphi}{4x}\right) - \left.\left(1 + \frac{2}{\epsilon x}\right)^{-3/4}\sin\left(\varphi + \frac{3\varphi}{4x}\right)\right], \quad (6.45)$$

which, together with (6.21)–(6.23), defines the density of states $v(E)$ in parametric form. Note the existence of scaling: if the energy is measured in units of $\Gamma_c$, and the density of states in $\Gamma_c/|\tilde{u}|$, then all the dependences are expressed in terms of universal functions uninfluenced by the degree of disorder. At $|E| \gg \Gamma$, we have the asymptotic relations

$$v(E)$$
$$= \begin{cases} \frac{1}{2}K_4 E^{(d-2)/2}\left[1 - \left(\frac{E}{\Gamma_c}\right)^{-\epsilon/2}\right]^{-1/4}, & E \gg \Gamma, \\ \frac{\Gamma_0(E)}{4\pi|\tilde{u}|}\left\{1 - \frac{R_0^\epsilon}{2}\left(\frac{|E|}{\Gamma_c}\right)^{-\epsilon/2} - \left[1 - \left(\frac{|E|}{\Gamma_c}\right)^{-\epsilon/2}\right]^{1/2}\right\}, & -E \gg \Gamma, \end{cases}$$
$$(6.46)$$

which exhibit a spurious pole. With large positive $E$, the function $v(E)$ gives the density of states of an ideal system; with large negative $E$ we get the result for the fluctuation tail

$$v(E) = \frac{K_4}{\pi} \Gamma_0(E) |E|^{-\epsilon/2} \ln \frac{1}{R_0}$$

$$= \tilde{c}_2 K_4 \left(\frac{2\pi}{3} \ln \frac{1}{R_0}\right)^{1/2} R_0^{-3} |E|^{(d-2)/2} \left(\frac{\bar{I}_4 |E|^{\epsilon/2}}{4|u|}\right)^{(d+1)/2}$$

$$\times \exp\left[\frac{2K_d I_4(R_0)}{\epsilon} - \frac{I_4(R_0)|E|^{\epsilon/2}}{4|u|R_0^\epsilon}\right], \qquad (6.47)$$

which corresponds to the traditional version of the instanton method. This result can also be obtained from Eqn (6.39). The divergence at $\epsilon \to 0$ is removed for a finite cutoff parameter $\Lambda$. It is interesting that for $\epsilon x \ll 1$, expressions (6.21) – (6.23), (6.45) functionally coincide with those for $d=4$ [38]; in other words, the behavior of all physical variables in the neighborhood of the mobility edge is effectively four-dimensional. The point of the phase transition moves to the complex plane, which ensures regularity of the density of states for all energies.

The fact that $R_0^\epsilon$ is different from unity is only important for $\varkappa^\epsilon/u \gg 1/\epsilon$, when the terms in braces in Eqn (6.46) cancel out almost exactly. The replacement $R_0 \to 1$ in Eqn (6.41) corresponds to a total neglect of $[\Sigma(p,\varkappa)]_{\text{nonpert}}$, since the integration in Eqn (6.44) is carried out in the range $p \gtrsim \varkappa$. Thus, the quantity $[\Sigma(p,\varkappa)]_{\text{nonpert}}$ is only important for large negative $|E|$, which justifies its calculation directly by Lipatov's asymptotics.

## 7. Perspectives of construction of a complete theory of the Anderson transition: the role of $\epsilon$ expansion

As follows from Section 1, the calculation of the density of states and the calculation of the conductivity of a disordered system, which are determined, respectively, by the mean Green's function $\langle G(x,x') \rangle$ and the correlator $\langle G^R G^A \rangle$, are two essentially different problems. Because of this, the values of the upper critical dimension for these two problems could in principle have been different. Such an affirmation was explicitly made in Ref. [32], which, however, is flawed with serious mistakes [36]. In reality, this is not the case: as follows from the arguments developed above, the special role of the dimensionality $d=4$ is most fundamentally manifested through the renormalizability of the theory, and the situation with renormalizability is the same for both problems. The latter follows if only from the fact that the diagram techniques employed in both cases are the same [13, 14]. Accordingly, the upper critical dimensionality for the problem of conductivity is four as well (see also the argument with the Ioffe – Regel criterion in Section 2), and the feasibility of $\epsilon$ expansion in the neighborhood of this dimensionality is certain.

The special role of the dimensionality $d=4$ for the problem of conductivity is clearly visible in the symmetry approach to the calculation of the critical exponents proposed by the author [19]. This approach opens up the possibility of constructing a complete theory of the Anderson transition, in which the $\epsilon$ expansion should play a very special role.

The 'symmetry theory' [19] is based on the physical idea of the relation between the phenomenon of localization and the diffusion pole in the irreducible four-leg vertex

$$U_{\mathbf{kk}'}(\mathbf{q}) = U_{\mathbf{kk}'}^{\text{reg}}(\mathbf{q}) + U_{\mathbf{kk}'}^{\text{sing}}(\mathbf{q})$$

$$= U_{\mathbf{kk}'}^{\text{reg}}(\mathbf{q}) + \frac{F(\mathbf{k},\mathbf{k}',\mathbf{q})}{-\mathrm{i}\omega + D(\omega,\mathbf{k}+\mathbf{k}')(\mathbf{k}+\mathbf{k}')^2}, \qquad (7.1)$$

put forward by Vollhardt and Wölfle in their so-called 'self-consistent localization theory' [21, 65, 66]. This idea agrees with the assumptions of the theory of weak localization [67, 68], according to which the diffusion pole in $U_{\mathbf{kk}'}(\mathbf{q})$ determines the main quantum corrections to the conductivity, which in turn determine the scaling behavior in a space of dimensionality $d = 2 + \epsilon$. The diffusion pole in $U_{\mathbf{kk}'}(\mathbf{q})$ with the 'classical' diffusion coefficient $D_{\text{cl}}$ results from summation of the 'fan' diagrams [68]. Vollhardt and Wölfle assumed that if *all* diagrams are taken into account, then $D_{\text{cl}}$ is replaced by the exact coefficient of diffusion $D(\omega,\mathbf{q})$. In the quantum kinetic equation, the quantity $U_{\mathbf{kk}'}(\mathbf{q})$ acts as the 'probability of transition', and using an estimation similar to the $\tau$ approximation, $D \propto l \propto \langle U \rangle^{-1}$ (where $l$ is the free path, and $\langle \ldots \rangle$ denotes averaging with respect to momenta), it is easy to obtain the self-consistency equation of the Vollhardt – Wölfle theory,

$$D \sim \text{const} \times \left(U_0 + F_0 \int \frac{\mathrm{d}^d q}{-\mathrm{i}\omega + D(\omega,q)q^2}\right)^{-1}, \qquad (7.2)$$

which in the original paper [21] was derived by crude solution of the Bethe – Salpeter equation. As the disorder increases, the 'probability of transition' grows catastrophically owing to the decrease of the diffusion coefficient, which ensures the possibility of its vanishing. Neglecting the spatial dispersion of $D(\omega,\mathbf{q})$, Eqn (7.2) allows the critical exponents of the conductivity $\sigma$ and localization radius $\xi$ to be found, which turn out to be equal to

$$s = 1, \quad d > 2; \quad v = \begin{cases} \dfrac{1}{d-2}, & 2 < d < 4, \\ \dfrac{1}{2}, & d > 4. \end{cases} \qquad (7.3)$$

These values agree miraculously with most of the known results, which makes one suspect that they are exact [69].

Analysis reveals [19] that a number of relations in the Vollhardt – Wölfle theory, initially obtained in Ref. [21] under speculative assumptions, can be rigorously proved. In particular, it is possible to prove the existence of the diffusion pole in $U_{\mathbf{kk}'}(\mathbf{q})$ [with the observed diffusion coefficient $D(\omega,\mathbf{q})$], the result for $D(\omega,\mathbf{q})$ in the localized phase

$$D(\omega,\mathbf{q}) = (-\mathrm{i}\omega) d(q), \quad \omega \to 0 \qquad (7.4)$$

and the relation between the coefficient of diffusion and the radius of localization of the wave functions, $D(\omega,0) \sim (-\mathrm{i}\omega)\xi^2$.

These results have led to a highly acute statement of the problem of the Anderson transition. Indeed, from Eqn (7.4) we see that $D(0,\mathbf{q}) \equiv 0$ in the localized phase; then the question arises concerning the behavior of the spatial dispersion of $D$ in the neighborhood of the transition. The most natural answer is that $D(0,\mathbf{q})$ becomes zero at the point of transition simultaneously for all $\mathbf{q}$. This hypothesis was put forward by Efetov [33]; the same assumption was used by Vollhardt and Wölfle. In the context of phenomenological reasoning in the spirit of Landau's theory, such a possibility is wholly unlikely: indeed, the entire function must vanish simultaneously irrespective of the way the critical point is approached and of the location of the point on the critical surface. Obviously, this cannot happen by accident, and must be backed by some profound symmetry. Does such a symmetry exist, and what kind of symmetry is it? In other words, such a scenario implies that the order parameter is a function rather than a number; the impact of this conclusion on the structure of the theory is clear. Another possibility is that $D(0, \mathbf{q})$ becomes zero at some point, after which the instability develops leading to a first-order phase transition. Then one has to suggest an appropriate scenario. Without answers to these questions one cannot speak of understanding the Anderson transition.

This problem is aggravated to the utmost by the existence of Ward's identity (1.12). The left-hand side of Eqn (1.12) is regular at the point of transition, whereas at $\omega \to 0$ the integrand on the right-hand side diverges in the localized phase for all $\mathbf{k}, \mathbf{k}'$ [see Eqns (7.1), (7.4)]. This singularity ought to be canceled after integration with respect to $\mathbf{k}'$, which involves $D(\omega, \mathbf{k} + \mathbf{k}')$ and imposes stringent requirements on the approximation used for calculating the spatial dispersion of the diffusion coefficient.

How are these problems solved in the existing theories? Currently regarded as the most rigorous is the approach to the theory of localization based on the $\sigma$-model formalism [70–72], which is derived by using the saddle-point approximation with respect to the 'hard' directions in the functional integral, and subsequent expansion in gradients. The 'minimal' $\sigma$-model is confined to the lowest (second) powers of gradients, which amounts to neglecting the spatial dispersion of the diffusion coefficient — the problems are ignored rather than resolved. The inclusion of spatial dispersion of $D(\omega, \mathbf{q})$ (inclusion of higher gradients into the Lagrangian of the $\sigma$ model) leads to a calamity ('the high-gradients catastrophe'): these terms grow rapidly in the course of renormalization-group transformations [73, 74]. The Vollhardt–Wölfle approach in this respect is more advanced since it leads to a physically clear-cut statement of the problem. In the original version [21, 65], however, the spatial dispersion of $D(\omega, \mathbf{q})$ was disregarded, and Ward's identity was crudely violated.

The symmetry theory [19] is based on analyzing the spectrum of the quantum collisions operator $\hat{L}$, which is a symmetrized version of the integral operator arising in the Bethe–Salpeter equation (1.11) as a result of Ward's identity (1.12)

$$[-\omega + (\epsilon_{\mathbf{k}+\mathbf{q}/2} - \epsilon_{\mathbf{k}-\mathbf{q}/2})]\phi_{\mathbf{k}\mathbf{k}'}(\mathbf{q})$$
$$+ \frac{1}{N}\sum_{\mathbf{k}_1} U_{\mathbf{k}\mathbf{k}'}(\mathbf{q})[\Delta G_{\mathbf{k}_1}(\mathbf{q})\phi_{\mathbf{k}\mathbf{k}'}(\mathbf{q}) - \Delta G_{\mathbf{k}}(\mathbf{q})\phi_{\mathbf{k}_1\mathbf{k}'}(\mathbf{q})]$$
$$= \Delta G_{\mathbf{k}}(\mathbf{q})N\delta_{\mathbf{k}-\mathbf{k}'}. \qquad (7.5)$$

The presence of a diffusion pole in the kernel of the operator $\hat{L}$ [see Eqn (7.1)] and the fact that $D(\omega, \mathbf{q}) \sim \omega$ lead in the localized phase to the decomposition

$$\hat{L} = \hat{L}_{\text{reg}} + \hat{L}_{\text{sing}} = \hat{L}_{\text{reg}} + \frac{\hat{L}_1}{\omega}, \qquad (7.6)$$

where in the operator $\hat{L}_1$ we have passed to the limit $\omega \to 0$ (the higher-order terms in $\omega$ are included in $\hat{L}_{\text{reg}}$). If the eigenvalue of $\hat{L}_1$ is finite, then it corresponds to the eigenvalue $1/\omega$ of the operator $\hat{L}_{\text{sing}}$, and, changing little upon addition of $\hat{L}_{\text{reg}} \sim 1$, generates the eigenvalue $1/\omega$ of the total operator $\hat{L}$. The zero eigenvalues of the operator $\hat{L}_1$ correspond to the zero eigenvalues of $\hat{L}_{\text{sing}}$, which after addition of $\hat{L}_{\text{reg}}$ become of the order of 1; some of them, however, turn out to be proportional to $\omega$ (Fig. 12). The latter follows from the fact that one of the eigenvalues $\lambda_0$ of the operator $\hat{L}$ differs only by a trivial factor from the diffusion coefficient $D(\omega, \mathbf{q})$. The invariance with respect to time reversal have a consequence that, along with $\lambda_0$, an infinite number of eigenvalues exhibit the same property. Hence, it follows that the number of zero eigenvalues of the operator $\hat{L}_{\text{sing}}$ is infinite.

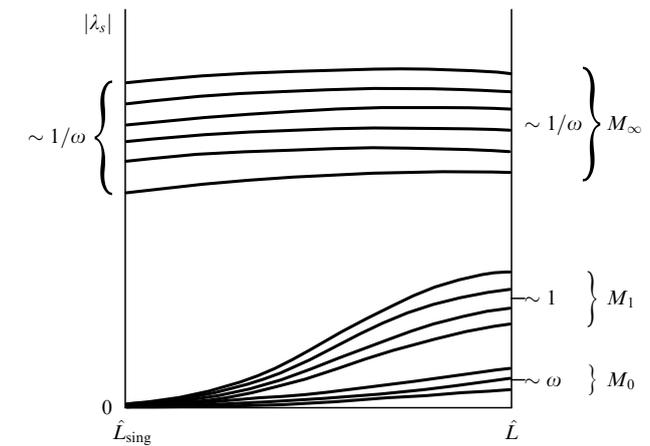

**Figure 12.** Evolution of the spectrum of eigenvalues $\lambda_s$ upon transition from $\hat{L}_{\text{sing}}$ to $\hat{L}$, that is, upon 'gradual switching on' of the operator $\hat{L}_{\text{reg}}$.

Decomposition (7.6) is similar to the decomposition of the Hamiltonian into symmetrical and asymmetrical parts, performed in the symmetry analysis of various physical problems: the operator $\hat{L}$ is represented as the sum of the operator $\hat{L}_{\text{sing}}$ of a higher symmetry (which is manifested in high number of zero modes), and the regular operator $\hat{L}_{\text{reg}}$ of the general form. Let us consider the response of the system to a perturbation $\delta\hat{L}_{\text{reg}}$: this gives rise to a problem of stability of the set $M_0$ of eigenvalues $\lambda_s \sim \omega$ of operator $\hat{L}$. Assume that the system occurs deep in the localized phase. Then a small perturbation $\delta\hat{L}_{\text{reg}}$ does not remove it from the state of localization, and the proportionality $\lambda_s \sim \omega$ is preserved for $s \in M_0$. On the other hand, if the perturbation $\delta\hat{L}_{\text{reg}}$ is of a general form, then it has nonzero matrix elements with respect to the eigenvectors of the operator $\hat{L}$, and must lead to values of $\lambda_s$ which are small but do not vanish for $\omega \to 0$.

This contradiction can be resolved in the following fashion. Operator $\hat{L}_{\text{reg}}$ acts over the complete Hilbert space $\Omega$, whereas the operator $\hat{L}_1$ has nonzero eigenvalues $\sim 1$ only in the subspace $\Omega_1$ which is part of $\Omega$ ($\Omega = \Omega_0 \oplus \Omega_1$). The change in $\delta\hat{L}_{\text{reg}}$ causes 'rotation' of the subspace $\Omega_1$, which gives rise to an effective disturbance $\delta\hat{L}_{\text{eff}}$ in the space $\Omega_0$ which compensates the initial perturbation $\delta\hat{L}_{\text{reg}}$. The condition of such a compensation leads to a self-consistency equation which replaces the crude Vollhardt–Wölfle equation. A solution of the self-consistency equation is sought under the assumption of arbitrary spatial dispersion of the coefficient of diffusion $D(\omega, \mathbf{q})$; however, only the solution with a weak dependence on $\mathbf{q}$ is intrinsically consistent, which does not affect the evaluation of the integral in Eqn (7.2) and leads to the result (7.3) for the critical exponents. In this way, all the main results of Ref. [21] are true, which is surprising for such a crude theory.

The statement of the absence of spatial dispersion of $D(\omega, q)$ on the scale $q \sim \xi^{-1}$ in the limit of $\omega \to 0$ sharply contradicts the earlier results of Refs [75, 76], according to which $D(\omega, q) \sim q^{d-2}$. In fact, the latter result can be disproved [77] on the basis of Eqn (7.4), which is a direct implication of the Berezinskiĭ–Gor'kov criterion [19]. Recently the results of Refs [75, 76] have been criticized in the context of the multifractal approach [78], which relies on the Chalker hypothesis [79], which holds that $D(\omega, q) \sim \omega^{\eta/d} q^{d-2-\eta}$ at the point of transition, where the anomalous exponent $\eta$ is surely different from zero and is related to the effective fractal dimension $D_2$ of the wave functions ($\eta = d - D_2$). The statement made in Ref. [19] that $D(0, q)$ becomes zero simultaneously for all $q$ is consistent with the Chalker hypothesis, while the assumption of the absence of spatial dispersion indicates that $\eta = d - 2$. The last conclusion is, however, premature: the results of Ref. [19] hold in the limit of $\omega \to 0$, and they can be 'pushed' to the smallest characteristic scale $\omega_0$ defined by the condition $\xi \sim L_{\omega_0} \propto \omega_0^{-1/d}$ [78], which vanishes at the point of transition. The result $\eta = d - 2$ can only be obtained by matching the results for $\xi \gg L_\omega$ and $\xi \ll L_\omega$, which still needs to be validated†. At the same time, the arguments against the equation $\eta = d - 2$ do not seem essential to us. Recent numerical calculations give $\eta = 1.2 \pm 0.15$, $\eta = 1.3 \pm 0.2$, $\eta = 1.5 \pm 0.3$ for three different methods [78], and do not exclude the possibility of $\eta = 1$ for $d = 3$. The result $\eta = 2\epsilon$ in the $2 + \epsilon$ theory, which is obtained by comparison with Wegner's work [81], seems to be inconsistent: the spatial dispersion is neglected in the derivation of the Lagrangian of the $\sigma$ model, to arise unexpectedly at some later point. The common argument that the fractal dimension can take on any value and does not need to be a 'good' number can also be refuted: the condition that the transition point corresponds to a fixed point of the renormalization group imposes certain restrictions on the nature of multifractality, which may be expressed by the equation $\eta = d - 2$. Note finally that Refs [75–79] use the relation between the density correlator $S(\omega, q)$ and the coefficient of diffusion,

$$S(\omega, q) \sim \frac{D(\omega, q) q^2}{\omega^2 + [D(\omega, q) q^2]^2}, \qquad (7.7)$$

which only holds if the coefficient of diffusion $D(\omega, q)$ is real, while the general relation is

$$S(\omega, q) \sim \frac{1}{\omega} \operatorname{Im} \frac{D(\omega, q) q^2}{-\mathrm{i}\omega + D(\omega, q) q^2} \qquad (7.8)$$

(see Eqn (31) in Ref. [19]). In the lowest order in $\omega$, the diffusion coefficient is real in the metallic phase and purely imaginary in the localized phase. In the neighborhood of transition it undergoes a sophisticated rearrangement, which is not even mentioned in the above references.

---

† The nontrivial nature of such matching is clear even from the fact that the coefficient of diffusion is complex-valued; there are, however, even more definite indications [80]. Accordingly, the result for $\tau \lesssim \omega^{1/(2\nu+1)}$ in Eqn (116a) in Ref. [19] is somewhat conditional, since the higher-order terms in $\omega$, which are small when $\tau \neq 0$, may become important at the point of transition itself.

The theory presented in Ref. [19] is strict under the assumption that the only singularities are the diffusion poles, whose existence is demonstrated from general principles. Other kinds of singularities may exist in specially designed models, but must be absent in the general case being not backed by the symmetry. Such reasoning is typical of the mean-field theory, and may be wrong because of some hidden elements of symmetry. There are indications, however, that this theory is something more than just the mean field theory.

As a matter of fact, the existence of hidden elements of symmetry is characteristic only of the critical point itself. Accordingly, in the typical case, the mean-field theory does not give the true critical behavior, but correctly describes the *change of symmetry*. According to the proposed scenario, the Anderson transition in terms of the change of symmetry is similar to the Curie point for an isotropic $n$-component ferromagnet in the limit of $n \to \infty$: while the change of magnetic field in the ferromagnet leads to rotation of the magnetization vector, the change of $\hat{L}_{\mathrm{reg}}$ in our case results in rotation of the infinite-dimensional subspace $\Omega_1$. The model of a ferromagnet with an infinite number of components is the basis of the $1/n$ expansion [9]; its critical exponents are known exactly and are in precise agreement with the results of straightforward analysis (7.3). This is an argument in favor of complete elucidation of the symmetry of the critical point and the exact determination of the exponents. The isotropy of the equivalent ferromagnet is the symmetry that ensures the simultaneous vanishing of $D(0, \mathbf{q})$ for all $\mathbf{q}$.

Further evidence is that the values of exponents (7.3) fall in agreement with all the reliable results of the model calculations (this was the rationale for the assumption that the results (7.3) are exact, which was expressed in Ref. [69]):

(a) For $2 < d < 4$, the exponents $s$ and $\nu$ comply with the Wegner relation $s = \nu(d - 2)$, which follows from the existence of one-parameter scaling [67], and for $d = 2 + \epsilon$, we have the results

$$\nu = \frac{1}{\epsilon}, \qquad s = 1 \qquad (\epsilon \to 0), \qquad (7.9)$$

which follow from the regular expansion for the Gell-Mann–Low function [67]

$$\beta_{\mathrm{GL}}(g) = \epsilon + \frac{A}{g} + \frac{B}{g^2} + \ldots \qquad (7.10)$$

with $A < 0$. The term $A/g$ in Eqn (7.10) at $d = 2$ defines the logarithmic corrections to the conductivity, and its existence (with the correct sign for $A$) can be checked by the diagram technique. In the 'minimal' $\sigma$ model, the first two corrections in $\epsilon$ to Eqn (7.9) vanish, while Wegner's third-order correction [82] impairs the agreement with numerical simulations and is questioned by the author himself. It looks like the equivalence of the null-component $\sigma$ model with the original disordered system takes place in the lowest orders in $\epsilon$. However it is, the problem of the high gradients calls for modifying the $\sigma$ models. Result (7.9), however, will survive all modifications which do not challenge the general philosophy of one-parameter scaling.

(b) Result (7.3) singles out the dimensionalities $d_{\mathrm{c}1} = 2$ and $d_{\mathrm{c}2} = 4$, which (from other independent considerations) are the lower and upper critical dimensionalities.

(c) All our experience with the theory of phase transitions indicates that for $d > d_{\mathrm{c}2}$, the critical exponents do not depend on $d$, which is true of Eqn (7.3).

(d) At $d = \infty$, the value of the exponent $v = 1/2$ agrees with the results for the Bethe lattice [83–85]; for $s$ we have two competing results, $s = \infty$ [85] and $s = 1$ [86], one of which agrees with Eqn (7.3).

(e) The overall behavior of $v$ as a function of $d$ is supported by estimates based on hierarchical models [18].

(f) The value of $v = 1$ for $d = 3$ is in good enough agreement with the results of early numerical calculations ($v = 1.2 \pm 0.3$ [87], $v = 0.9 \pm 0.3$, $v = 1.4 \pm 0.2$), but the recent trend is toward somewhat higher values: $v = 1.35 \pm 0.15$ [89], $v = 1.54 \pm 0.08$ [90], $v = 1.45 \pm 0.08$ [91]. We believe, however, that it is too early to consider this trend in earnest. The higher accuracy in the recent works has been achieved through more sophisticated processing, whereas the 'raw' data, as admitted by the authors of Ref. [89], remain at the benchmark of 1990 [88], and are not expected to change much in the near future [88]. The reported accuracy does not reflect the systematic error, which is certainly large in systems of small size $L$, with $L/a_0 \leqslant 13$ [89, 90]. The same data, with crude corrections to the scaling, can be reconciled with a value of $v = 1$ [92, 93]. What is more, the method of evaluation of the correlation length $\xi$ based on the Lyapunov index for the quasi-one-dimensional system of size $M \times M \times L$ with $L \gg M$, used by most authors [87–90], does not have a solid theoretical background. In our opinion, all points complying with the condition $\xi \gtrsim M$ should be excluded, while nothing is currently done now to discriminate them. Among these works, the study of Ref. [91] stands out: it is based on a straightforward diagonalization of the Hamiltonian and considers systems of a relatively large size $L^3$ with $L/a_0 = 28$; however, it also takes into consideration points with $\xi \gtrsim L$. Note finally that the empirical formula for $v$ vs. $d$, obtained in Ref. [35], contradicts the theoretical result $v = 1/\epsilon$ for $\epsilon = d - 2 \to 0$, which unambiguously follows from the one-parameter scaling (the base of all treatment in Ref. [35]), and the result $d_{c2} = 4$, which, in our opinion, is convincingly validated in the present review.

In this way, the theory of Ref. [19] may claim that it gives the exact values of the critical exponents. To complete the theory, one must prove that there are no singularities other than the diffusion poles. The general mathematical proof seems improbable, since it is always possible to construct a model with a 'built-in' singularity. The more natural approach would therefore consist in checking the *general structure* of the theory [19] against a physically reasonable model. Such test for the $(4 - \epsilon)$-dimensional Gaussian model (1.1) is quite feasible. If successful, it will symbolize not only the construction of an $\epsilon$ expansion, but the complete solution of the problem of the Anderson transition.

The author is grateful to M V Sadovskiĭ for reading the early version of the manuscript and making valuable comments, V E Kravtsov and M V Feĭgel'man for stimulating discussions, and members of the seminars at the Institute for Physical Problems (IFP) and the Physical Institute of the Academy of Sciences (FIAN) for their interest and advice.

This work was financially supported by INTAS (Grant 96-580) and the Russian Foundation for Basic Research (RFFI Project 96-02-19527).